\documentclass[letterpaper,twocolumn,10pt]{article}
\usepackage{usenix-2020-09}
\usepackage{authblk} 
\usepackage{tikz}
\usepackage{amsmath}
\usepackage{pifont}
\usepackage{multirow}
\usepackage{comment}

\usepackage{graphicx}  
\usepackage{float}  
\usepackage{subcaption}

\usepackage{filecontents}

\newcommand{\quan}[1]{\textcolor{red}{Quan: #1}}
\newcommand{\SJC}[1]{\textcolor{purple}{SJC: #1}}

\newcommand{\sysname}{{\textsc{FastLibra}}}

\author{Hang Zhang\textsuperscript{*}, Jiuchen Shi\textsuperscript{*}, Yixiao Wang\textsuperscript{}, Quan Chen\textsuperscript{$\dagger$}, Yizhou Shan\textsuperscript{}, Minyi Guo\textsuperscript{}}
\affil{\textit{\normalsize{\textsuperscript{}Shanghai Jiao Tong University}}}

\begin{document}

\date{}



\title{Improving the Serving Performance of Multi-LoRA Large Language Models via Efficient LoRA and KV Cache Management}

\maketitle


\begin{abstract}
Multiple Low-Rank Adapters (Multi-LoRAs) are gaining popularity for task-specific LLM applications.
For multi-LoRA serving, caching hot KV caches and LoRA adapters in high bandwidth memory of accelerations can improve inference performance.
However, existing Multi-LoRA inference systems fail to optimize serving performance like Time-To-First-Toke (TTFT), neglecting usage dependencies when caching LoRAs and KVs. 
We therefore propose \textbf{\sysname{}}, a Multi-LoRA inference caching system to optimize the serving performance. \sysname{} comprises a \textit{dependency-aware cache manager} and a \textit{performance-driven cache swapper}.
The cache manager maintains the usage dependencies between LoRAs and KV caches during the inference with a unified caching pool.
The cache swapper determines the swap-in or out of LoRAs and KV caches based on a unified cost model, when the HBM is idle or busy, respectively.
Experimental results show that \sysname{} reduces the TTFT by 63.4\% on average, compared to state-of-the-art works.
\end{abstract}

\renewcommand{\thefootnote}{\fnsymbol{footnote}}
\footnotetext[1]{Hang Zhang and Jiuchen Shi contributed equally to this work.}
\footnotetext[2]{Quan Chen is the corresponding author.}

\section{Introduction}

Large Language Models (LLMs) are now widely used to understand and generate human-like text~\cite{mann2020language,chowdhery2023palm}. 
While it is cost-inefficient to train LLMs for different tasks,  parameter-efficient fine-tuning~\cite{hu2021lora,wang2024lora} that freeze the large-scale base model and 
fine tunes multiple Low-Rank Adapters (LoRAs) for different tasks are increasingly popular~\cite{dettmers2024qlora,alpacalora}. 
For instance, in applications like chatbot~\cite{ChatGPT,Bard}, personal agents~\cite{apple-assistant,li2024personal}, and multi-language translation~\cite{zhang2016google,zhang2020improving},
multiple LoRAs can be tuned for different user languages and application scenarios. 
For these LLMs, Key-Value (KV) caches that store input context are often used to maintain coherence and speed up responses during extended interactions by avoiding repetitive computations~\cite{kwon2023efficient,agrawal2023sarathi}.
Researchers also proposed to reuse history KVs for queries with the same prefix~\cite{zheng2023efficiently,gim2024prompt,yao2024cacheblend,yu2023stateful}, boosting performance in iterative tasks.

\begin{figure}
    \centering
    \includegraphics[width=.9\linewidth]{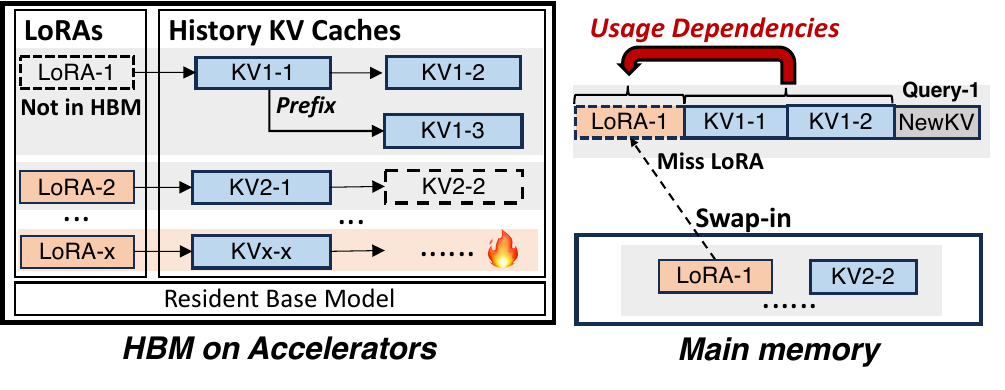}
    \caption{An example caching state to show the \textbf{\textit{Usage Dependencies}} between LoRAs and KV caches.}
    \label{fig:intro}
   \vspace{-3mm}
\end{figure}

To improve the serving performance of such Multi-LoRA applications, 
many works have investigated to cache the base model, the KVs~\cite{yu2023stateful,gao2024attentionstore,qin2024mooncake} or ``hot'' LoRA adapters (LoRAs in short)~\cite{sheng2024slora,iliakopoulou2024chameleon,li2024caraserve}, in the high bandwidth memory (HBM) of accelerators (e.g., global memory of Nvidia GPUs or global memory of various AI accelerators~\cite{hbm-wikipedia,nvidia-a100,google-tpu-intro}). 
While caching both KVs and LoRAs can improve inference performance, vLLM~\cite{kwon2023efficient} and SGLang~\cite{zheng2023efficiently} proposed to cache both LoRAs and KVs.
\autoref{fig:intro} shows an example of caching base model, LoRAs, and KVs.
In general, LoRAs have separate KV caches (e.g., LoRA-1 and LoRA-2). 
Moreover, the HBM space is statically partitioned for caching LoRAs and KVs,
because these works allocate different sizes of memory blocks for LoRAs and KVs, preventing their sharing with each other~\cite{vLLM2024code}.

When a user query is received, the serving system checks whether the required LoRA and KVs are already in the HBM or not. 
If the required LoRAs and/or KVs are not cached, they are swapped in from the main memory. If the cache space for the LoRAs/KVs is full, some of them 
are swapped out with various caching policies.
When queries use different LoRAs following stable distributions, this solution performs well because the optimal HBM space partition can be identified in a ``brute-force'' way.
However, production traces~\cite{shahrad2020serverless,zhang2021faster,zheng2023judging} show that the distributions are dynamic. 
In such scenario, we observe that static HBM partition and independent cache management suffer from low efficiencies in both intra-LoRA and inter-LoRA aspects. 

In the intra-LoRA aspect, a query's KVs may remain cached while its required LoRA is swapped out.
As shown in \autoref{fig:intro}, when ``Query-1'' relying on LoRA-1 arrives, it can run only if LoRA-1 and its prefixed KV1-1 and KV1-2 are in HBM.
However, LoRA-1 is swapped out earlier due to the limited HBM space.
In this case, the cached KVs are actually ``invalid'', because the query cannot run without the required LoRA, showing their inherent \textit{usage dependencies}.
If the HBM space of invalid KVs (e.g., KV1-3) were used to cache LoRA-1, Query-1 could run immediately.
Invalid KVs of a LoRA may also prevent useful KVs of other LoRAs from being cached. 
For instance, KV2-2 is not cached while LoRA-1's KVs are invalid, preventing queries of LoRA-2 from running.
Our experiments show that vLLM~\cite{kwon2023efficient} suffers from up to 46.5\% invalid KV caches.

In the inter-LoRA aspect, the required number of LoRAs and the hotness of KVs for different LoRAs change dynamically, due to the varying loads of different LoRAs.
For the example in \autoref{fig:intro}, more LoRAs (e.g., LoRA-x) need to be used at the next time interval, and the KVs of LoRA-x become hot but other LoRAs' KVs have occupied the HBM which prevents them to be swapped-in.
However, with static HBM partition of LoRAs and KVs, their swap-in or out can only be managed separately, making it hard to uniformly balance the usage of LoRAs and KVs in HBM.
To address the above problems, 
a scheme is required to integrate the usage dependency for each LoRA and its KVs 
with a unified caching pool.
By such means, we can keep valid KVs in HBM as much as possible to improve the HBM efficiency.
Moreover, unified managing the 
caching of LoRAs and KVs helps in reducing the Time-To-First-Token (TTFT) and Time-Per-Output-Token (TPOT) of each query. %
It is challenging to 
to balance the usage of LoRAs and KVs. 

Based on the two insights, we propose \textbf{\sysname{}}, a Multi-LoRA inference caching system that optimizes the caching of LoRAs and KVs considering the usage dependency.
\sysname{} aims to reduce the TTFT, while maximizing the peak serving throughput of Multi-LoRA applications.
It comprises a \textit{dependency-aware cache manager} and a \textit{performance-driven cache swapper}.
The cache manager maintains the usage dependencies between KV caches and LoRAs based on a tree-based scheme with a unified caching pool, where nodes are KV caches or LoRAs and edges are their dependencies. 
To maintain usage dependencies during inference, LoRAs or KV caches are inserted or removed from leaves in the HBM to keep the tree connected.
Based on the unified caching management of the cache manager, the cache swapper periodically
determines the swap-in/out of LoRAs and KVs by using
a unified cost model that precisely reflects the benefits of swap-in/out LoRAs and KVs to the performance of queries. This paper makes three contributions.

\begin{itemize}
    \item \textbf{Investigating the caching management of LoRAs and KVs for the Multi-LoRA inference.} The analysis motivates us to maintain the usage dependencies between LoRAs and KV caches, and unified manage their swap-in/out based on their impact on inference performance. 
        \item \textbf{The design of a scheme that maintains the usage dependencies between LoRAs and KV caches.} Considering the usage dependencies, 
 more valid KVs are cached in HBM, eliminating the intra-LoRA inefficiency.
    \item \textbf{The design of a cost model that guides the swap-in/out of LoRAs and KVs.} 
    The model enables the unified swap-in and out of LoRAs and KVs, eliminating the inefficiency due to the inter-LoRA interference.
\end{itemize}


We have implemented \sysname{} on top of vLLM~\cite{martinez2024impact} and evaluated it with Llama-7B/13B/34B models~\cite{touvron2023llama} on four High-performance NPUs with three typical scenarios (chatbot~\cite{luo2024arena,chiang2024chatbot}, multi-language translation~\cite{zhang2020improving}, and personal agents~\cite{byrne-etal-2019-taskmaster}). 
The design of \sysname{} does not rely on any specific hardware architecture, and it is applicable to other accelerators.
Experimental results show that \sysname{} reduces the TTFT and TPOT by 63.4\% and 40.1\%, respectively, as well as improves the peak serving throughput by 35.2\%, compared to state-of-the-art Multi-LoRA inference systems.


\section{Background and Motivation}
In this section, we first introduce the background of Multi-LoRA serving with caching, then investigate the inefficiency of current Multi-LoRA serving systems.

\subsection{Multi-LoRA Serving with Caching}


\textbf{Multi-LoRA.} 
LoRA is a popular method for efficiently fine-tuning pre-trained LLMs by adding lightweight adapters to original weights~\cite{hu2021lora}. 
Instead of updating all parameters of a model, LoRA only learns a pair of small low-rank matrices that modify the original weights.
These matrices are much smaller than the original weight matrix, which can reduce the computational cost and memory usage.

For the Multi-LoRA scenario, the pre-trained base model is loaded once, and multiple pairs of low-rank matrices are introduced, each corresponding to a specific task~\cite{huang2023lorahub,zhao2024lora}.
For each task $t$, a unique pair of low-rank matrices $A_{t}$ and $B_{t}$ is learned, and the original weight matrix $W$ is updated as:
\begin{equation}
    \small
    \label{eq:bg_eq1}
    \begin{aligned}
        W_t'=W+\Delta W_t=W+A_t B_t
    \end{aligned}
\end{equation}

For Multi-LoRA serving, based on the query's task, the corresponding LoRA matrices are loaded into the HBM for usage before inferencing.
Queries using different LoRAs can be processed in a single batch using Segmented Gather Matrix-Vector multiplication (SGMV)~\cite{sheng2024slora,chen2024punica}, improving both efficiency and throughput.

\textbf{KV Caches for Multi-LoRAs.} 
The LoRAs need to be loaded into the HBM for usage during the inference~\cite{hu2021lora,sheng2024slora}.
Moreover, most LLMs use a decoder-only transformer to predict the next token with KV caches computed from previous tokens~\cite{floridi2020gpt,chowdhery2023palm}. When a query matches an existing prefix, the stored KV caches are reused to reduce HBM usage and eliminate redundant computations, such as in multi-turn dialogues~\cite{gim2024prompt,gao2024attentionstore}. Therefore, maintaining the history KV caches in HBM can maximize the reuse.

Each LoRA adds a low-rank branch to the original weights that participate in the KV cache computation. For each query $q$ using LoRA $t$, the KV cache is computed as:
\begin{equation}
    \small
    \label{eq:bg_eq2}
    \begin{aligned}
        KV\_Cache_{q,t}=W_{k,v} q + A_t B_t q
    \end{aligned}
\end{equation}
Therefore, as mentioned in \autoref{fig:intro}, KV caches for different LoRAs are stored separately due to task-specific modifications by each LoRA branch~\cite{hu2021lora,dettmers2024qlora}.


The separate storage further increases contention for limited HBM space, and thus the KV caches and LoRAs are usually offloaded to main memory and swapped-in/out on-demand~\cite{gao2024attentionstore,sheng2024slora}.
However, this can cause cold-starts when loading them back into HBM, affecting performance metrics like TTFT and TPOT.
To reduce this overhead, we need to pre-cache ``hot'' history KV caches and LoRAs into HBM.

\textbf{Multi-LoRA Serving.} 
When a new query arrives, if the required LoRA is not in the HBM while the HBM is full, this query needs to queue to wait for other KV caches or LoRAs swapping-out from the HBM, and then loading the required LoRA.
Similarly, if the required KV caches are not in HBM, they will be swapped-in from the main memory.
Once the required LoRA and KV caches are properly loaded and matched, the inference processes to generate the next token. 

The LLM inference typically has the prefill and decode stages~\cite{yu2022orca,agrawal2023sarathi}, corresponding to two performance metrics of the Time to First Token (TTFT) and Time Per Output Token (TPOT). 
The above workflow can introduce overheads due to the queue to wait for HBM space, LoRA cold-starts, and KV cold-starts, affecting both TTFT and TPOT.

\subsection{Low Multi-LoRA Serving Performance\label{moti-1}}
We use 
vLLM~\cite{vLLM2024code} that caches both LoRAs and history KVs~\cite{hu2021lora,zheng2023efficiently} as the representative serving system to perform the investigation. 
It allocates fixed HBM space for LoRAs and KV caches, and utilizes Least Recent Use (LRU) policy 
to swap-in/out LoRAs or KV caches in the respective HBM area.
vLLM sets a predefined allocation ratio of the HBM space for the LoRA (empirically set to be 0.2) and the memory block size to 32, referring to the latest version of vLLM~\cite{vLLM2024code}. 


Three Multi-LoRA scenarios, {\it chatbot}, {\it multi-language translation}, and {\it personal agent} are used as benchmarks in the investigation. 
LMSYS-33k~\cite{zheng2023judging}, Opus-100~\cite{zhang2020improving}, and Taskmaster~\cite{byrne-etal-2019-taskmaster} datasets are used to generate the queries, respectively.
Since datasets Opus-100 and Taskmaster lack the timing information, the same to state-of-the-art Multi-LoRA management works~\cite{wu2024dlora,sheng2024slora,iliakopoulou2024chameleon}, we use Microsoft Azure Function Trace~\cite{shahrad2020serverless,zhang2021faster} to adapt its query arriving timing information.
Moreover, we use Llama-7B/13B/34B as the base models, and conduct the experiments on four High-performance NPUs.
\autoref{table:config} summarizes the hardware and software configurations.



\begin{table}
    \small
    \setlength\tabcolsep{2pt}
    \centering  
    \caption{Experiment specifications}
    \begin{tabular}{c|c}
      \hline
        & \textbf{Specifications} \\
      \hline
      \multirow{3}*{\textbf{Hardware}}
       & Arm CPU (192 cores), 256GB main memory \\
       &  NPU with 256 TFLOPS FP16 and 64GB HBM $\times$ 4\\
       &  PCIe $\times$ 16 Gen4.0\\
       \hline
       \multirow{3}*{\textbf{Software}} 
       & Llama-7B, Llama-13B, and Llama-34B\\
       & LMSYS-33K~\cite{zheng2023judging}, Opus-100~\cite{zhang2020improving}, Taskmaster~\cite{byrne-etal-2019-taskmaster},\\
       & Microsoft Azure Function trace~\cite{shahrad2020serverless,zhang2021faster}\\
        \hline 
    \end{tabular}
    \label{table:config} 
   \vspace{-3mm}
  \end{table}
\begin{figure}
    \centering
    \includegraphics[width=.81\linewidth]{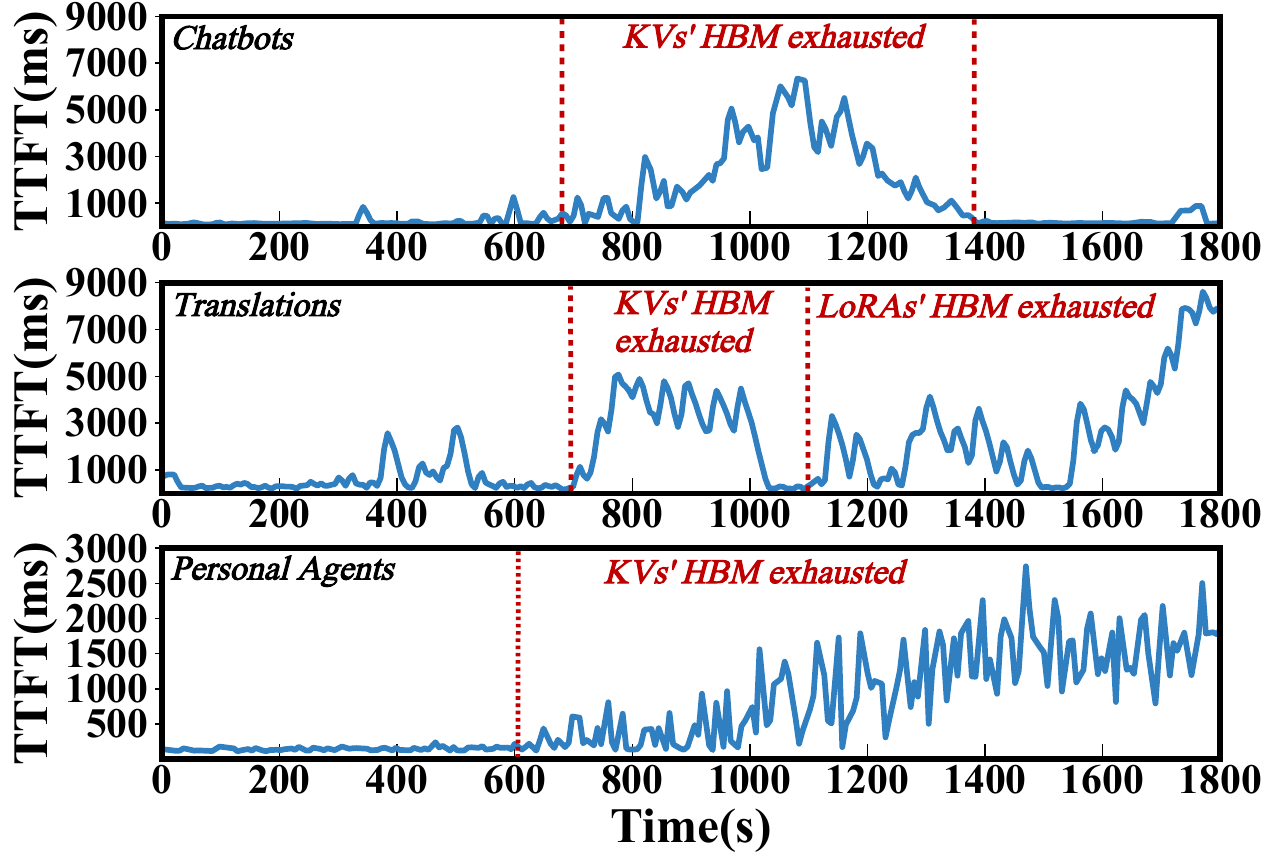}
    \caption{The TTFT of vLLM for various scenarios.}  
    \label{fig:moti_vllm}
   \vspace{-3mm}
\end{figure}
\begin{figure*}
    \centering
    \includegraphics[width=.95\linewidth]{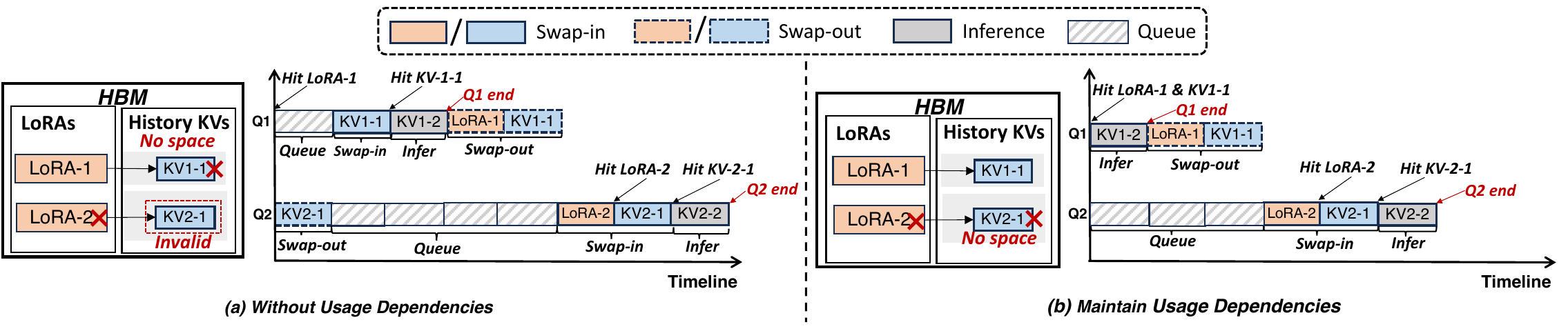}
    \caption{Examples of serving two queries under: (a) without usage dependencies, and (b) maintaining usage dependencies.}
    \label{fig:moti_example}
   \vspace{-3mm}
\end{figure*}

\autoref{fig:moti_vllm} shows the TTFT of vLLM for the three benchmarks 
with the Llama-7B base model. Experiments with other base models show similar observations, as shown in \autoref{eval}.
With varying loads, we observe that vLLM experiences significantly high TTFT at certain periods, due to insufficient HBM space allocation for KV caches or LoRAs.
As statistics, the TTFT of the three benchmarks are 1032.4ms, 1905.1ms, and 730.8ms on average, respectively.
This is because the static HBM allocation of vLLM cannot dynamically adapt to the varying loads in Multi-LoRA serving. 
The HBM allocation is static because vLLM allocates memory blocks with different sizes for LoRAs and KVs according to their respective requirements~\cite{vLLM2024code}.
Memory blocks in the KV cache HBM area cannot be used for LoRAs, and vice versa, making it impossible to dynamically adjust the pool sizes.

While redeployment can change the HBM partition, it results in significant overhead that blocks the normal inference for tens of seconds~\cite{aryan2023costly,alizadeh2023llm}. 
Moreover, even if dynamic HBM allocation is achieved with more fine-grained memory blocks~\cite{chen2024punica,sheng2024slora,iliakopoulou2024chameleon}, 
it is still challenging to define an appropriate allocation policy with varying loads of LoRA adapters.




\subsection{Diving into Underlying Reasons\label{moti-2}}
Our investigations show that the poor serving performance 
is caused by 1) \textit{inefficient HBM usage without considering intra-LoRA usage dependencies}, 
and 2) \textit{inappropriate swap-in/out of KVs and LoRAs when LoRAs have varying loads}.

\subsubsection{Inefficient HBM Usage\label{moti-dependency}}

\autoref{fig:moti_example} shows an example of serving two queries ({\it Q1} and {\it Q2}) of two LoRA adapters ({\it LoRA-1} and {\it LoRA-2}).

As shown in \autoref{fig:moti_example}(a), it is possible that KV2-1 is cached while the corresponding LoRA-2 is not in the HBM, 
without considering usage dependencies between LoRA and its KVs. 
Prior work (e.g., vLLM, and SGLang) all manage LoRAs and KVs in this way.
In this case, KV2-1 is ``invalid'', since {\it Q2} cannot run at all without the LoRA adapter.
At the same time, $Q1$ is also blocked although its LoRA adapter is cached, because it needs to wait for the required KV1-1 to be swapped in, before which KV2-1 should be swapped out to free some HBM space.
After $Q1$ returns, $Q2$ needs to swap-in the LoRA-2 and KV2-1 again to perform the inference.
Without considering the usage dependency, the serving system causes redundant swap-in/out, greatly increasing the queuing overhead.
From our evaluations in \autoref{eval}, vLLM results in 48.1\% invalid KV caches on average. 

\autoref{fig:moti_example}(b) shows a better caching case where LoRAs and KVs are managed based on the usage dependency.
In this case, $Q1$ runs directly because both {\it LoRA-1} and the history KV1-1 are in the HBM. 
After $Q1$ returns, Q2 runs after LoRA-1 and KV1-1 are swapped out and the required LoRA-2 and KV2-1 are swapped in. 
In this way, the redundant swap-in/out is eliminated, and the response time of both {\it Q1} and {\it Q2} reduces.


{\it While prior work does not consider the usage dependency between LoRA and its KVs, the limited HBM space is not efficiently used.}

\subsubsection{Inappropriate Swap-in/out of KVs and LoRAs\label{moti-2-2}}
Previous works~\cite{vLLM2024code,zheng2023efficiently} separately manage LoRAs and KVs in individual HBM areas, and adopt caching strategies like Least-Recent-Used (LRU) for swap-in/out.
These works cannot dynamically balance the HBM usage for LoRAs and KVs when the loads of different LoRAs change.

\begin{figure}
    \centering
    \includegraphics[width=0.95\linewidth]{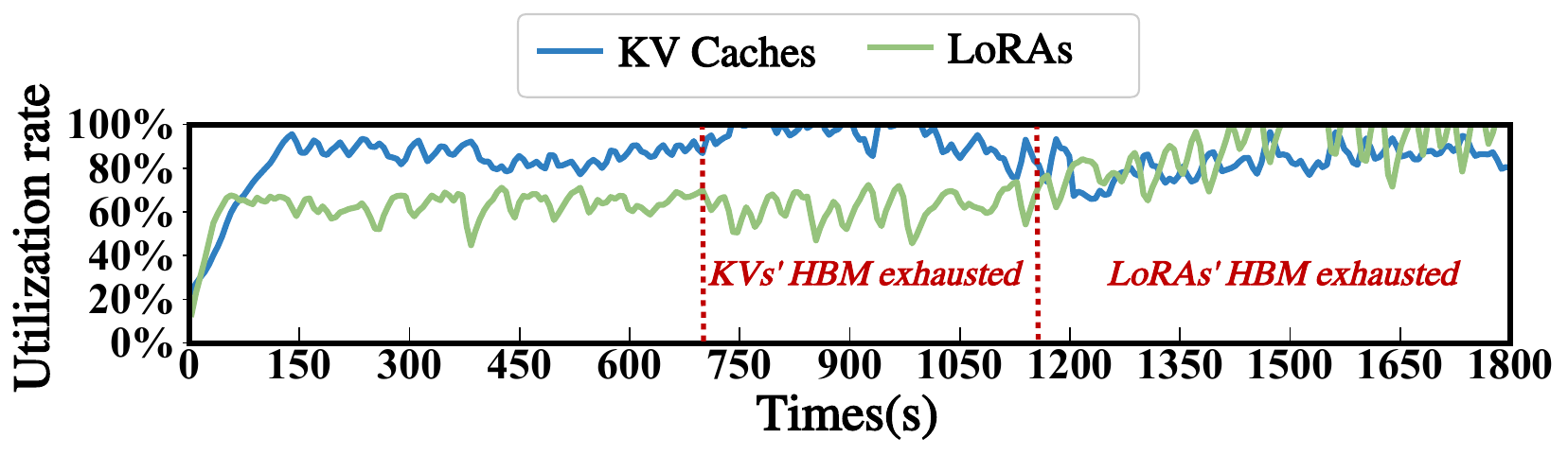}
    \caption{The utilization rate of the HBM space allocated to LoRAs and KV caches over time in the translation scenario.}
    \label{fig:moti_hbm_transaltion}
   \vspace{-3mm}
\end{figure}

Take the benchmark {\it translation} in \autoref{fig:moti_vllm} as an example. The TTFT increases up to 5036.1ms and 8617.9ms during the period of 700s-1100s and 1100s-1800s, respectively.
Correspondingly, \autoref{fig:moti_hbm_transaltion} shows the HBM utilization rates of the LoRA and the KV parts.
After looking into the detailed serving trace, we find that the long TTFT originates from different reasons.
During 700s and 1100s, the HBM space for KVs is exhausted while the utilization rate of HBM space for LoRAs is 58.9\% on average. 
In this case, the frequent swap-in/out of KVs results in the long TTFT. 
During 1100s and 1800s, the HBM space for LoRAs is exhausted on the contrary, 
because queries of more LoRA adapters are received during that period. 
According to the trace, queries of 41 LoRAs are received before 1100s, while that is 75 after 1100s.

\begin{figure}
    \centering
    \includegraphics[width=0.32\linewidth]{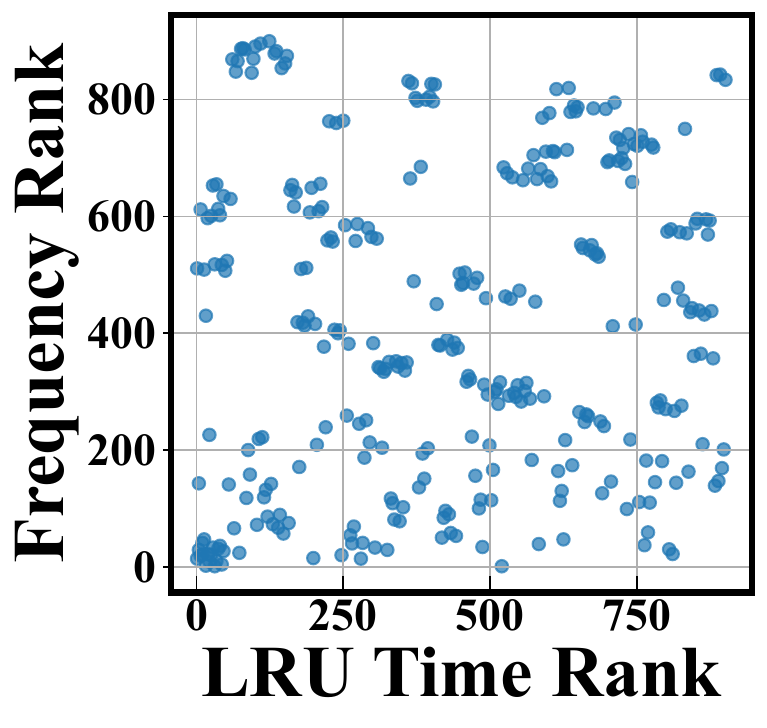}
    \includegraphics[width=0.32\linewidth]{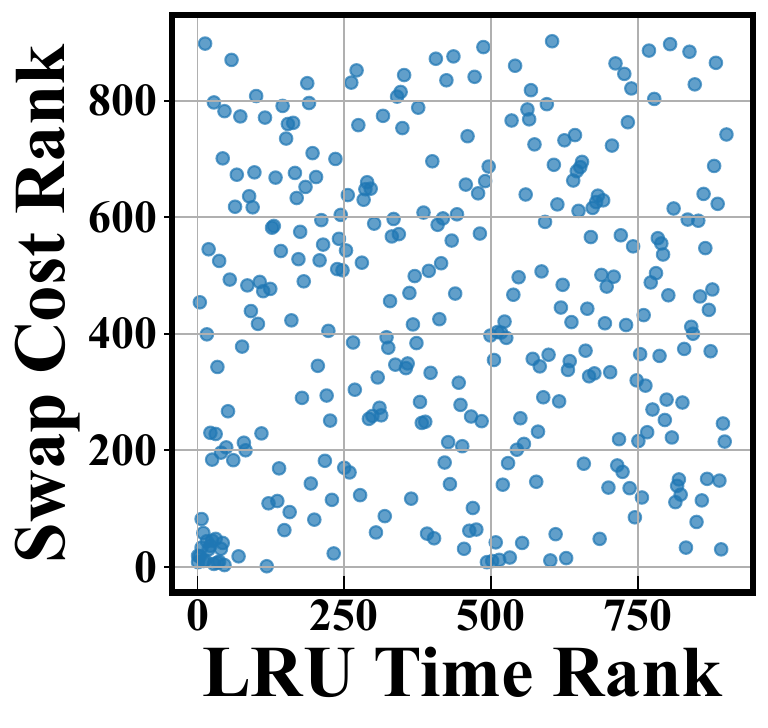}
    \includegraphics[width=0.32\linewidth]{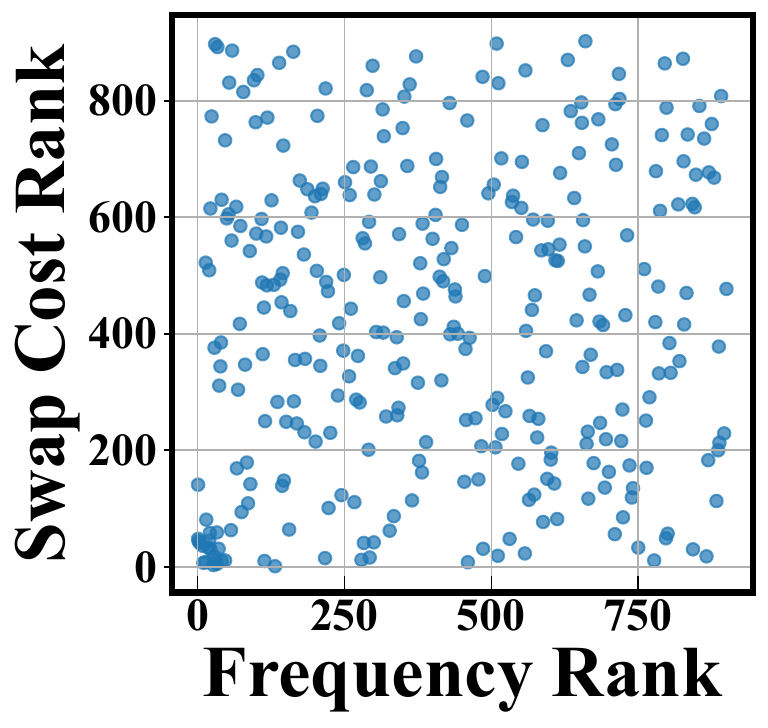}
    \caption{The relationships among the visited frequency, swap costs, LRU of the KV caches or LoRAs.}
    \label{fig:moti_LRU}
   \vspace{-3mm}
\end{figure}

It would be beneficial to dynamically balance the HBM usage of LoRAs and KVs.
However, even if the HBM space of LoRAs and KVs is dynamically managed through fine-grained memory blocks, 
relying on the LRU policy to determine the swap-in/out is not efficient. 
This is because the TTFT is related to many factors, like the swap cost and visited frequency.
\autoref{fig:moti_LRU} shows the relationship between the visited frequency, LRU time, and swap cost of each KV cache and LoRA.
In the figure, each point represents a LoRA or KV cache, and its $x$-axis or $y$-axis represents its corresponding ranks of LRU Time/Frequency/Swap Cost.
As observed, the points are randomly distributed, which means there is not clear correlation among these key factors.

{\it Relying on LRU to manage the HBM space is not efficient to minimize the TTFT, even if dynamic HBM usage is enabled.}

\section{\sysname{} Methodology}
In this section, we summarize the challenges of \sysname{}, and introduce the overview of \sysname{}.
\subsection{Challenges of \sysname{}}
According to the above analysis, two technical challenges should be addressed to resolve the above problems.

Firstly, the neglect of the intra-LoRA usage dependencies between LoRAs and KV caches brings invalid KV caches in HBM. This can prevent other useful KVs or LoRAs from being loaded, increasing the cold-start overheads. 
\textbf{A suitable scheme is required to construct the usage dependencies among the LoRAs and KV caches and consistently maintain the dependencies when serving the queries.}

Secondly, when loads of different LoRAs vary, the required HBM space for caching LoRAs and the hotness of KV caches of different LoRAs change accordingly.
Thus, we need to balance the HBM usage for LoRAs and KVs, and swap-in or swap-out the appropriate KVs and LoRAs to optimize the TTFT. 
\textbf{An appropriate mechanism is needed to access the benefits or harms of swapping-in or out each LoRA or KV cache to the TTFT of future queries.}

\begin{figure}
    \centering
    \includegraphics[width=1.0\linewidth]{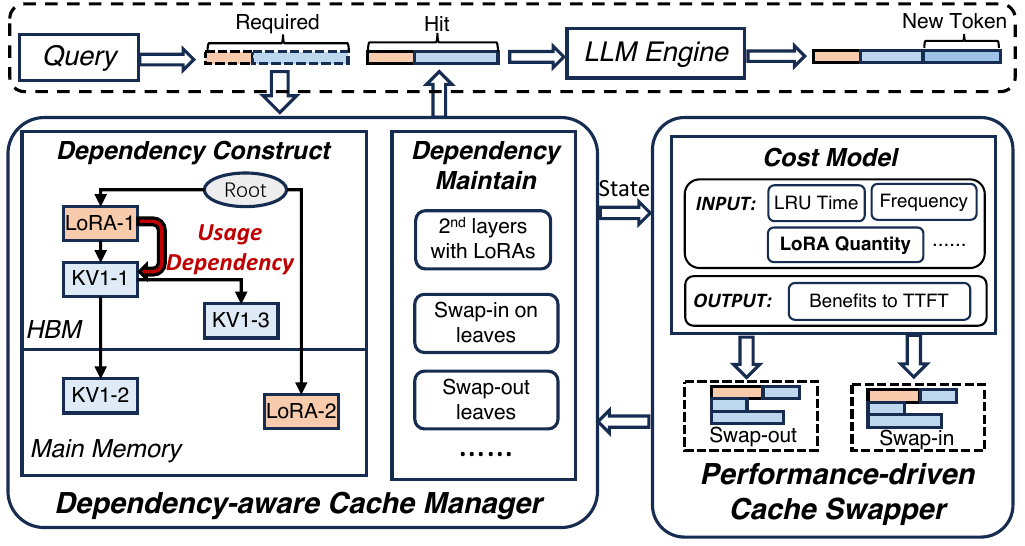}
    \caption{Design overview of \sysname{}.}
    \label{fig:overview}
    \vspace{-3mm}
\end{figure}

\subsection{Overview of \sysname{}}



\autoref{fig:overview} shows the design overview of \sysname{}. 
It comprises a \textit{dependency-aware cache manager} and a \textit{performance-driven cache swapper}.
The cache manager 
manages LoRAs and KV caches in the HBM and main memory together, and maintains usage dependencies to eliminate invalid KV caches. 
After each monitor interval, the cache swapper 
decides the swap-in or swap-out  LoRAs and KV caches from the main memory or HBM based on a unified cost model. 
The cache manager then conducts specific swapping operations. 

The most challenging part is managing LoRAs and KV caches based on the usage dependencies to eliminate invalid KV caches. \sysname{} introduces the tree-based dependency maintenance scheme to address this problem.
In the dependency tree, nodes represent LoRAs or KV caches, and edges represent the usage dependencies among them.
When a query arrives, its required LoRAs and KVs in this tree are matched according to the Depth-First-Search (DFS).
To maintain the usage dependencies, 
this scheme places the LoRAs on the second layer, as well as only swaps-out leaf nodes in the HBM and swaps in root nodes in the main memory (\autoref{sec:manager}).


When the loads of queries using different LoRA branches change, the used LoRA number can increase and the hotness of some KV caches of some LoRAs changes. 
\sysname{} periodically decides the swap-in/out of different LoRAs and KV caches based on performance metrics like LRU time, visit frequency, the LoRA quantity, etc.
The challenging part here is to establish the cost model to directly evaluate the benefits or harms to the TTFT of swapping-in/out each LoRA or KV cache (\autoref{sec:swapper}).



Specifically, \sysname{} 
works as follows.
1) The cache manager organizes LoRAs and KV caches in HBM and main memory, constructing their usage dependencies into a dependency tree.
2) During inference, it inserts newly loaded LoRAs into the second layer of the tree, and inserts or deletes KV cache nodes at the leaves of their corresponding LoRA branches. 
3) After each monitor interval, the cache swapper retrieves the states of nodes from the cache manager, and decides the swapped-in/out KV caches and LoRAs when the HBM is idle/busy.
The decisions are made using a cost model that considers metrics like LRU time, visit frequency, and loaded LoRA quantity, and accesses their impact on TTFT.
4) The swap-in/out decisions are sent back to the cache manager for performing corresponding memory operations.
5) For a new query, if its LoRAs or KV caches are in main memory but HBM is full, the cache manager swaps out ``cold'' LoRAs or KVs based on the cache swapper's decisions, 
then swaps in the required ones.
6) This query proceeds for inference to generate the next token with the required LoRA and KVs.

\sysname{} can be adapted to other LLM inference engines~\cite{zheng2023efficiently,yu2022orca,agrawal2023sarathi} by replacing their memory management module with few modifications. It applies to LLMs based on decoder-only transformer~\cite{touvron2023llama,floridi2020gpt,chowdhery2023palm} that cover popular LLM practical scenarios. 
The design of \sysname{} does not rely on any specific hardware architecture, and it is applicable to other accelerators.

\section{Dependency-aware Cache Manager\label{sec:manager}}
In this section, we first analyze how to construct the usage dependencies among LoRAs and KV caches,  then introduce their maintenance during serving the queries.

\begin{figure}
    \centering
    \includegraphics[width=1.0\linewidth]{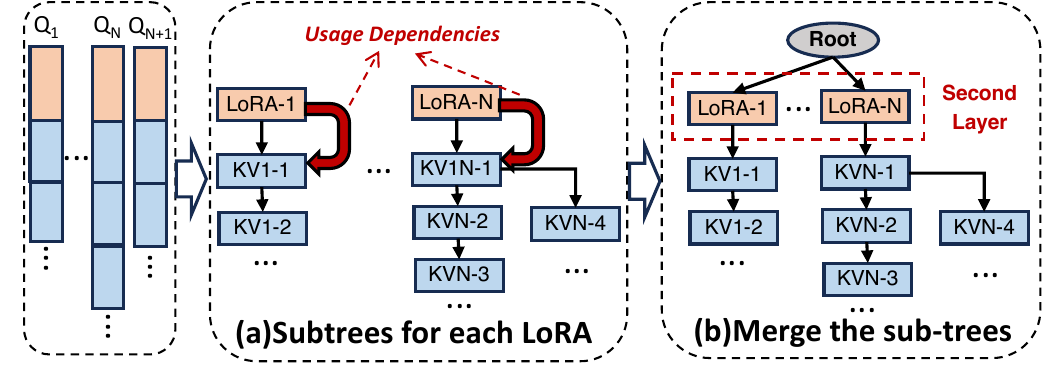}
    \caption{The constructing process of the usage dependencies among LoRAs and KV caches.}
    \label{fig:dependenecy_construst}
    \vspace{-3mm}
\end{figure}

\subsection{Usage Dependency Constructing}
As we analyzed in \autoref{moti-dependency}, a LoRA and its corresponding KV caches have their inherent usage dependencies. When ignoring these dependencies, invalid KV caches will occupy the HBM space, leading to low performance for Multi-LoRA inference.
In this subsection, we adopt a tree-based scheme to construct the usage dependencies among KV caches and LoRAs used by the queries in the HBM and main memory, as shown in \autoref{fig:dependenecy_construst}.

For each specific query, it will first match the required LoRA and then its corresponding KV caches. The KV caches corresponding to different tokens also have their matching orders. For instance, in the sentence ``To be or not to be'', the KV cache for the token ``To'' should be matched in front of ``be''. Therefore, as shown in \autoref{fig:dependenecy_construst}(a), the LoRAs and subsequent KVs can be intuitively connected by a chain like the branch of LoRA-1, where nodes represent LoRAs or KV caches and edges represent the usage dependencies.
Moreover, a KV cache for a token may have several possible subsequent KV caches. For instance, the subsequent tokens for the prefix sentence ``To be'' can be ``or not to be'' or ``the best''.
Thus, the LoRA and its subsequent KV caches can also construct a subtree like the branch of LoRA-N in this figure.

Since these subtrees constructed above are still separate, we need to merge these subtrees into a unified one, as shown in \autoref{fig:dependenecy_construst}(b).
We use a virtual root node to connect the subtrees for different LoRAs to form a unified tree. In this way, all LoRA nodes are placed on the second layer of the tree, and newly arrived queries can first match the required LoRA node in this tree.
Through the construction method described above, the usage dependencies among LoRAs and KV caches within the same LoRA branch is established, while different LoRA branches remain independent.

\subsection{Dependency Maintaining During Inference\label{dependency_maintain_sec}}

To maintain the usage dependencies among LoRAs and KV caches during query inference, we need to correctly match and update the LoRAs and KV caches on the dependency tree. Moreover, we need to swap-in and swap-out appropriate nodes in the dependency tree according to the cache swapper's decisions (\autoref{sec:swapper}) when the HBM is busy or idle.

\begin{figure}
    \centering
    \includegraphics[width=.9\linewidth]{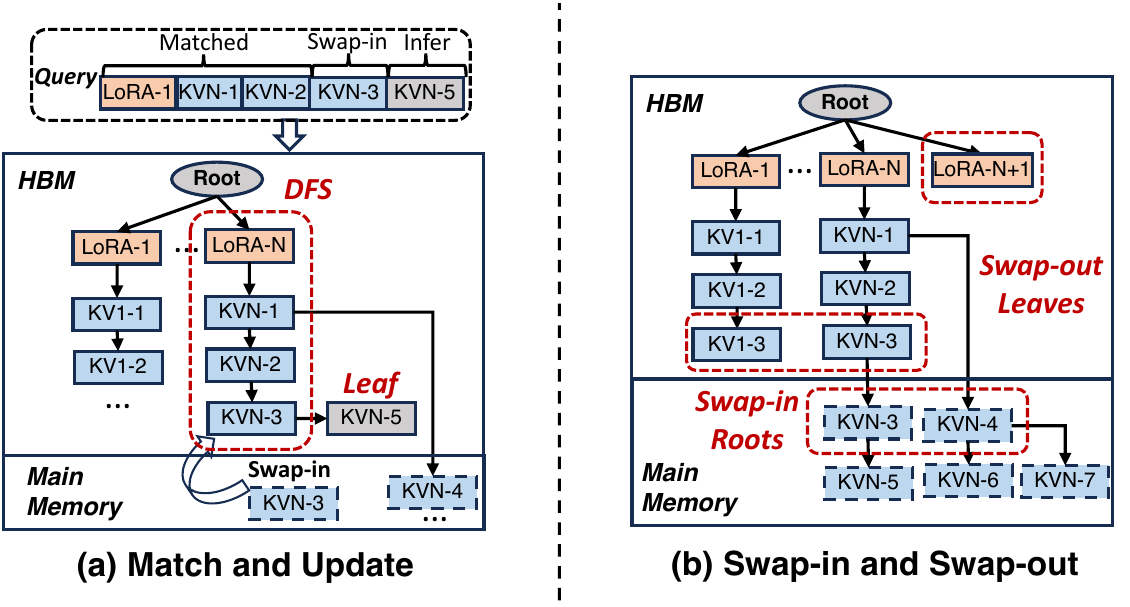}
    \caption{Maintaining the usage dependencies among LoRAs and KV caches during the query inference.}
    \label{fig:dependency_maintain}
    \vspace{-3mm}
\end{figure}

For the matching and updating, as shown in \autoref{fig:dependency_maintain}(a), when a query arrives, it needs to match the required LoRAs and KV caches.
This query will first match the LoRA node in the second layer. If the LoRA resides in the main memory, this node is swapped-in the HBM asynchronously. 
Then, within the subtree of this LoRA branch, this query begins to match history KV caches according to Deep-First-Search (DFS) of the tree until the leaf node is reached or no corresponding node can be found. 
During the KV matching process,
if the required KV cache resides in the main memory, it will first be swapped-into the HBM. 
Through the above prefix matching process, we can maximize the reuse of KV caches that have already been computed according to the usage dependencies.
At last, this query generates a new token with a new KV cache, and we will insert it below the last matched node of its corresponding LoRA subtree.
Also, during the decoding process, the new KV cache will continuously be inserted into the leaves of this LoRA branch.


When the HBM is idle or busy, some LoRAs or KV caches needs to be swapped-in or swapped-out to fully utilize the HBM and main memory resources.
As shown in \autoref{fig:dependency_maintain}(b), the cache manager will control the swapping-out to start from the leaf nodes in the HBM, as well as control the swapping-in to start from the root nodes of each subtree in the main memory.
This is because, during the node matching process in the dependency tree, the nodes higher up will always be prioritized for matching and all their children nodes depend on them.
In this way, the usage dependencies among LoRAs and KV caches can be maintained during the inference and all KV caches that reside in the HBM are valid ones, thus the HBM can be fully utilized.


\subsection{Implementing Unified Tree-based Caching}
\sysname{} is implemented based on vLLM~\cite{vLLM2024code} with an extra 8324 and 1644 lines of Python and C++ codes, respectively. We describe how \sysname{} establishes unified HBM and main memory pool for LoRAs and KVs, as well as implementing the usage dependency tree and asynchronous swap-in/out.

\textbf{Unified Caching Pool for LoRAs and KVs:}
To achieve a unified memory pool for HBM and main memory, we extend the BlockManager of vLLM~\cite{kwon2023efficient,zheng2023efficiently}. During the initialization phase, both HBM and main memory are partitioned into memory blocks of the same size. This block-wise memory allocation policy is similar to S-LoRA~\cite{sheng2024slora}, but we also extend this pool to store history KV caches. To retain LoRAs, we perform block-wise partitioning of LoRAs along the rank dimension. Since the other dimensions of LoRA align with those of the KV caches, this approach ensures full alignment with the KV cache and avoids memory fragmentation. 

When some LoRAs or KV caches are swapped-out, \sysname{} recycles their memory blocks in the memory pool of HBM for future allocations. Moreover, whenever a new KV cache is generated on HBM, it will be directly retained in the HBM without deciding to place it in the HBM or main memory. 
This eliminates redundant memory operations.

\textbf{Usage Dependency Tree:}
We build the usage dependency tree on top of the unified memory pool, which logically records the memory address of each memory block without altering the actual physical memory allocation pattern. We utilize an efficient trie tree~\cite{trie-wikipedia} to implement the usage dependency tree whose node matching and updating is fast as less than 1ms. In the usage dependency tree, each path from the root to a leaf represents a conversation record, and the subtrees with the same parent node have a shared prefix. 

The node label for each KV cache node is the token sequence, and for each LoRA node is the LoRA ID. Each node also retains the corresponding important information, i.e., visit frequency, last recent usage time, and the node size. These data will be updated when each node is generated, matched, or swapped-in/out.

\textbf{Asynchronous Swapping-in/out:} To further mitigate the cold start overhead, we adopted an asynchronous swap-in/out strategy similar to existing work~\cite{gao2024attentionstore}. We use the Stream library in Torch~\cite{torch-stream-docs} to implement this. After a query arrives, if the corresponding LoRA or KV caches for the query is not in HBM, we swap in the corresponding memory blocks and just let this query wait while inferring other requests that are ready. This realizes the overlap of inference and data transferring, thus improving the inference efficiency.



\section{Performance-driven Cache Swapper\label{sec:swapper}}
In this section, we first analyze the impact of the quantity of LoRAs loaded into HBM on TTFT. 
Then, we introduce a cost model considering multiple metrics to access the benefits to TTFT of swapping-in/out different LoRAs and KV caches.
At last, we introduce the workflow of the cache swapper.

\subsection{Considering LoRA Quantity on TTFT\label{LoRA_ensure}}
As the LoRA quantity used changes dynamically over time, the LoRA quantity in the HBM can impact the TTFT.

\begin{figure}
    \centering
    \includegraphics[width=.7\linewidth]{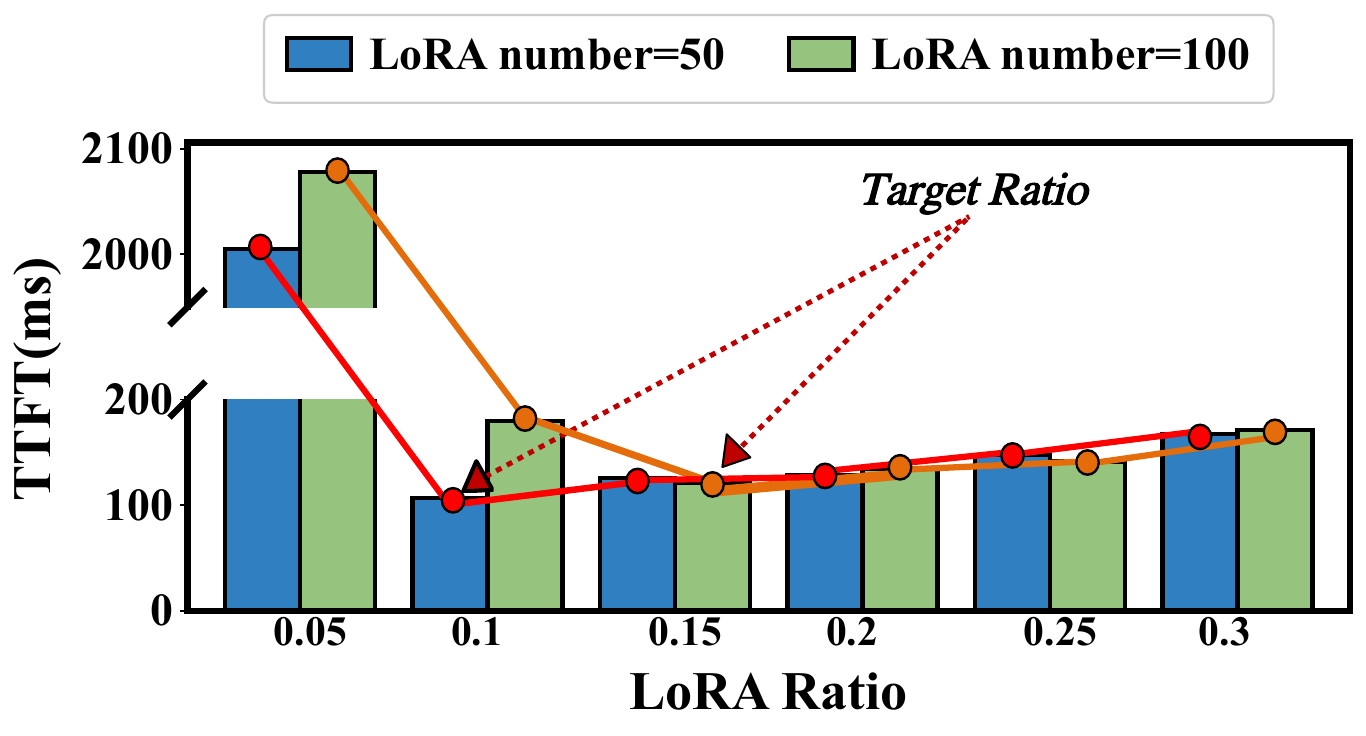}
    \caption{The TTFT of vLLM under different HBM allocation ratios for LoRAs.}
    \label{fig:moti_lora_minimum} 
    \vspace{-3mm}
\end{figure}

\autoref{fig:moti_lora_minimum} shows the TTFT under the chatbot scenario with different HBM allocation ratios for LoRAs in the vLLM. 
In this experiment, the used LoRA number is set at 50 and 100, as well as the average sending rate is 2 queries per second.
We can observe that the TTFT reduces significantly before reaching a target ratio, and the target ratio is increased when the required LoRA number changes from 50 to 100.
This is because the query inference can only start once the required LoRA is matched in HBM, otherwise the query is queued.
Insufficient LoRA loading quantity in HBM can cause a large amount of LoRA cold-starts, leading to a significant increase in TTFT.
Therefore, sufficient LoRA quantity is needed under different dynamic scenarios.



We estimate the current required LoRA quantity based on two factors: the usage frequency probability \( prob_i \) of LoRA \( i \), which is obtained from the recorded data in the dependency tree, and the recent inference batch size \( BS \) from the last 5 seconds. Using these, we calculate the expected number of LoRAs required for inference (\( Low_{lora} \)) as follows:
\begin{equation}
\small
    \begin{aligned}
        Low_{lora} = \sum_{i=1}^{n} fe_i = \sum_{i=1}^{n} \left(1 - (1 - prob_i)^{BS}\right)
    \end{aligned}
\label{eq:expected_batch_lora_num}
\end{equation}
In this formula, the $fe_i$ represents the probability that the LoRA $i$ is present in recent batch, i.e., 1 minus the probability that none of the queries in this batch use it.
We consider $Low_{lora}$ as an important aspect of cost model.










\subsection{Cost Model to Access Benefits to TTFT}
When performing swap-in or swap-out operations for LoRAs and KV caches, 
the goal is to retain the most valuable KVs and LoRAs in HBM as much as possible, thereby optimizing the TTFT for incoming queries. 
To achieve this, our key idea is to design a cost model to evaluate the expected benefits to TTFT of retaining a specific KV cache or LoRA $i$ in HBM. 

As analyzed in \autoref{moti-2-2} and \autoref{LoRA_ensure}, the cost model needs to try to load sufficient LoRAs, and consider metrics with the visited frequency, the LRU time, and the cost of swap-in/out of nodes. 
Thus, we first define the $LoRA\_Eva_i$ as the reward coefficient that encourages the loaded LoAR quantity to be close to the $Low_{lora}$ (\autoref{eq:expected_batch_lora_num}) as:
\begin{equation}
    \small
    \begin{aligned}
        LoRA\_Eval_i = & max(1, \frac{Low_{lora}}{Now_{LoRA}})
    \end{aligned}
\label{eq:lora_demand}
\end{equation}
In this formula, $LoRA\_Eval_i$ gives a larger reward when $Now_{LoRA}$ is farther from the $Low_{lora}$, and is set to 1 when $Now_{LoRA}$ is greater than or equal to $Low_{lora}$.

Then, we define the $Retain\_Eva_i$ to represent the expected benefit of retaining node $i$ in HBM, which can be estimated as the expected cold-start latency reduction to TTFT for future queries after caching it in HBM. The definition is as:
\begin{equation}
\small
Retain\_Eval_i = cost_i \times prob_i \times (1 - \text{sigmoid}(t_i))
\label{eq:reward_swapin}
\end{equation}
In this formula, the first item transfer $cost_i$ can be computed using the PCIe bandwidth and size of the KV or LoRA, and the second item visit frequency probability $prob_i$ is based on the recorded data on the dependency tree.
The third item is a time decay function similar to the forget-gates in the LSTM~\cite{staudemeyer2019understanding}, whose $t_i$ represents the time difference between the current time and LRU time. This function enhances the weight of KVs or LoRAs that were visited more recently.

Combining the formulas of the LoRA reward coefficient and the expected TTFT benefits of future queries, we finally design the cost model to access a KV cache or LoRA $i$ as:
\begin{equation}
    \small
    \begin{aligned}
        Eval_{i} = LoRA\_Eval_i  \times Retain\_Eval_i
    \end{aligned}
\label{eq:C_swap}
\end{equation}
As for the definition, a KV cache or LoRA with higher $Eval_i$ has more benefits to be stored in the HBM, and in other words, meaning it incurs higher costs if it is swapped-in HBM from the main memory.
This cost model evaluates the relative relationship between each KV cache or each LoRA in terms of benefits to the TTFT when retaining them in the HBM. Then, we can use these relationships to decide their swap-in and swap-out orders when the HBM is full or idle (\autoref{swapper_workflow}).

\begin{figure}
    \centering
    \includegraphics[width=.9\linewidth]{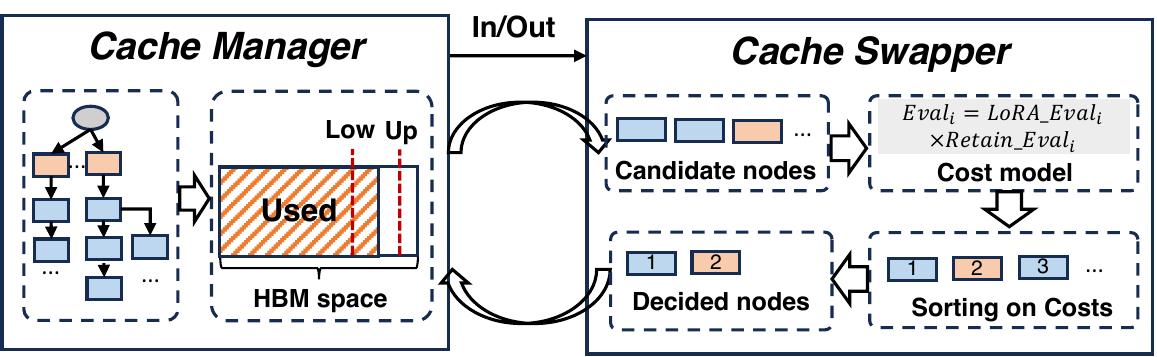}
    \caption{The operation workflow of the cache swapper.}
    \label{fig:swapper}
    \vspace{-3mm}
\end{figure}

\subsection{Workflow of the Cache Swapper\label{swapper_workflow}}
Based on our cost model in \autoref{eq:C_swap}, \autoref{fig:swapper} shows the operation workflow of the cache swapper under the cooperation with the cache manager.


After each monitor interval of 100ms, the cache manager first calculates the HBM usage based on the storage state of the usage dependency tree. We set HBM usage upper and lower thresholds (95\% and 70\% in our evaluations) to determine whether HBM is busy or idle, following existing cache management works~\cite{vLLM2024code,zhong2024distserve}. The upper threshold is set below 100\% to leave HBM space for KV caches generated by running queries.
Moreover, if we just set the upper threshold and directly swap-in/out according to it, HBM usage may be frequently higher/lower than it in a short time, leading to Ping-Pong swappings, and thus we set the lower threshold to address it.
If the HBM usage is larger than the upper threshold, the cache manager will send the swap-out instruction to the cache swapper. Moreover, according to \autoref{dependency_maintain_sec}, the leaf nodes in the HBM will be sent as candidate nodes to the cache swapper.
Similarly, if the HBM usage is smaller than the lower threshold, the swap-in instruction along with the root nodes of each path in the main memory will be sent. 

After receiving the candidate nodes from the cache manager, the cache swapper then assesses their benefits to inference performance based on the cost model in \autoref{eq:C_swap}.
If swap-in is currently required, the cache manager sorts the candidate nodes with the descending order of their $Eval_i$.
Otherwise, it sorts them with the increasing order for the swap-out. Following the greedy algorithm, the cache manager continuously swap-in or swap-out the nodes one by one according to the sorting until the HBM is at a balanced status.

\begin{figure*}[htbp]
    \centering
    \begin{subfigure}[b]{0.3\textwidth}
        \includegraphics[width=\textwidth]{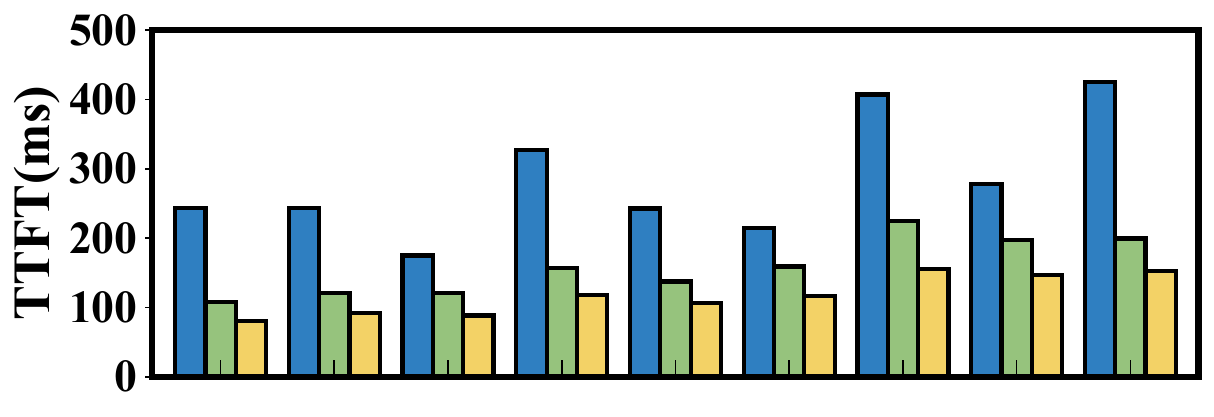}
    \end{subfigure}
    \begin{subfigure}[b]{0.3\textwidth}
        \includegraphics[width=\textwidth]{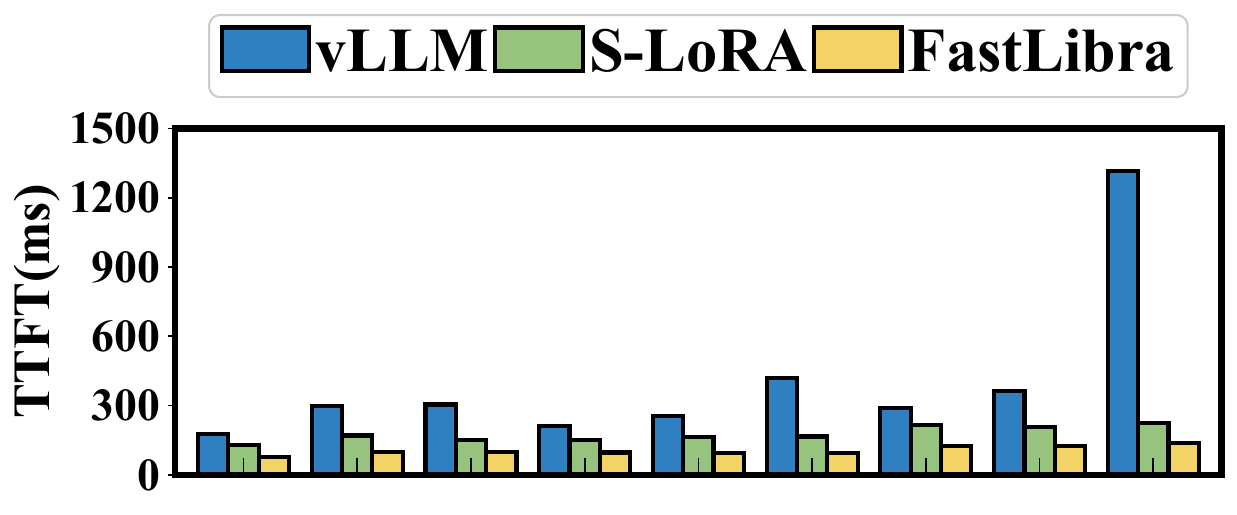}
    \end{subfigure}
    \begin{subfigure}[b]{0.3\textwidth}
        \includegraphics[width=\textwidth]{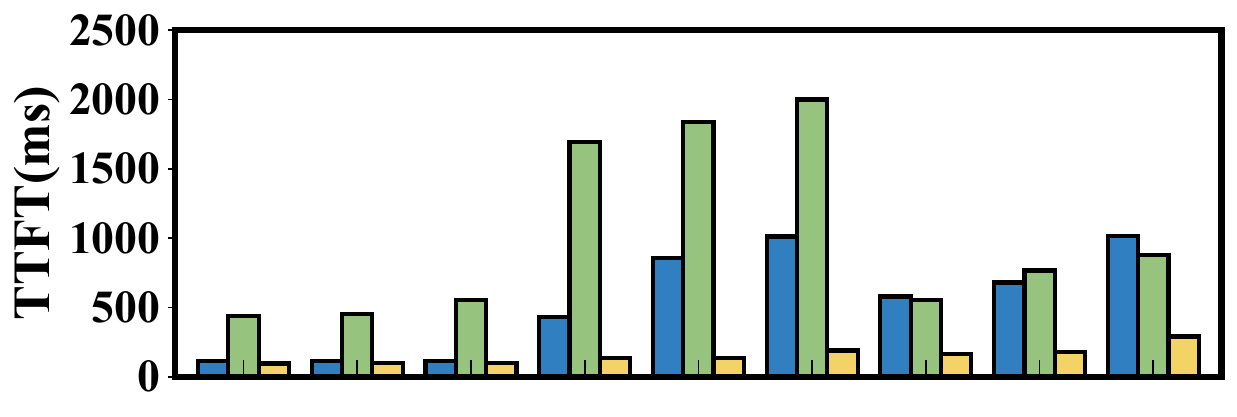}
    \end{subfigure}


    \begin{subfigure}[b]{0.3\textwidth}
        \includegraphics[width=\textwidth]{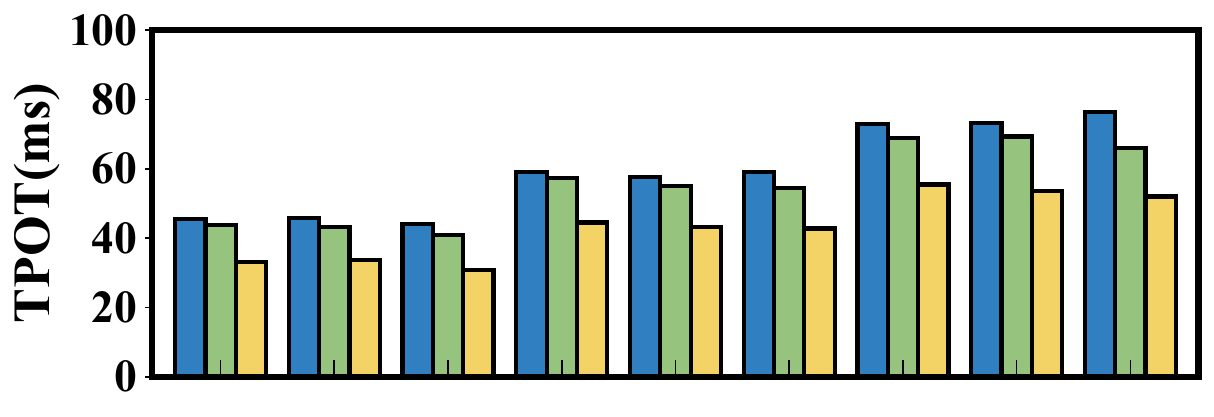}
    \end{subfigure}
    \begin{subfigure}[b]{0.3\textwidth}
        \includegraphics[width=\textwidth]{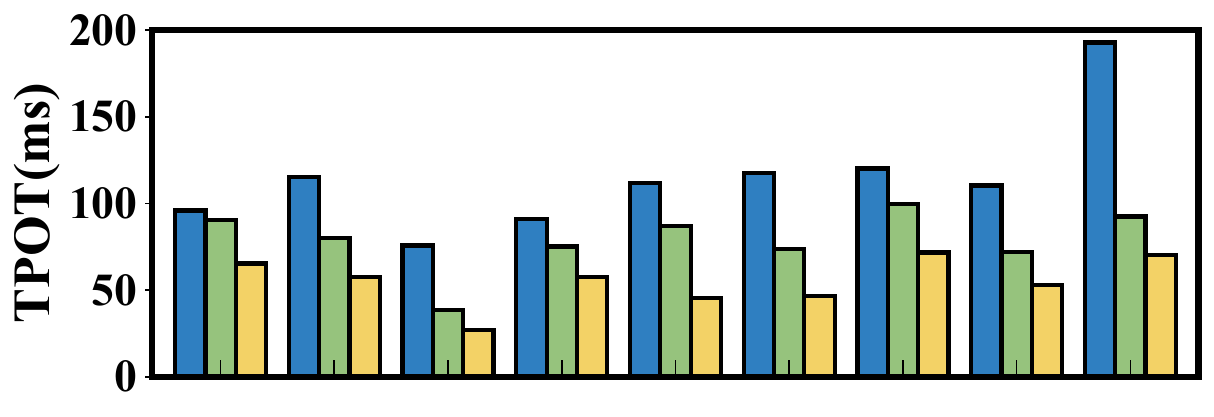}
    \end{subfigure}
    \begin{subfigure}[b]{0.3\textwidth}
        \includegraphics[width=\textwidth]{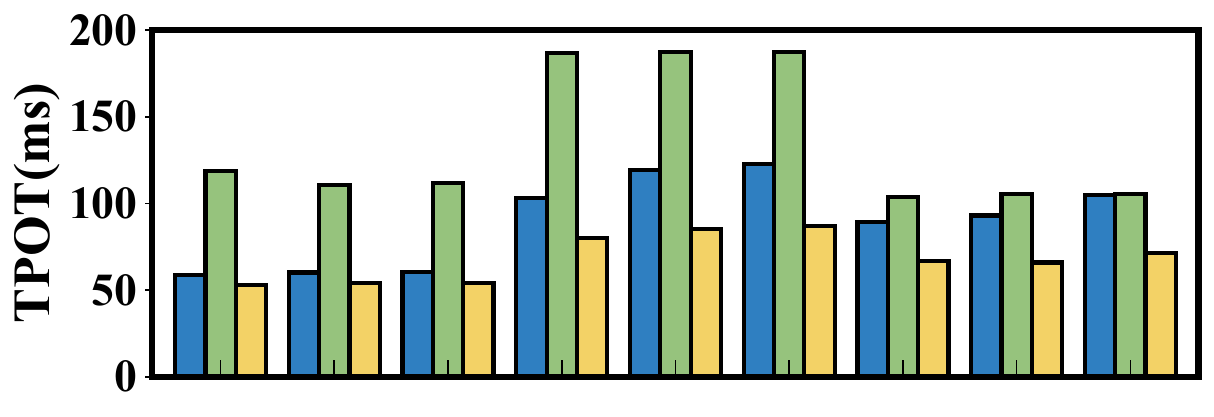}
    \end{subfigure}


    \begin{subfigure}[b]{0.3\textwidth}
        \includegraphics[width=\textwidth]{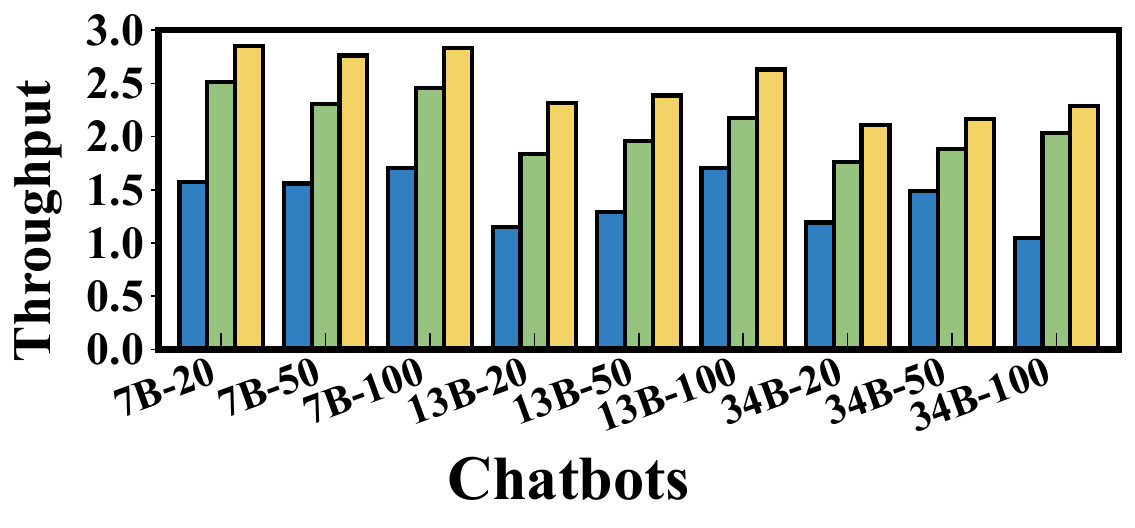}
    \end{subfigure}
    \begin{subfigure}[b]{0.3\textwidth}
        \includegraphics[width=\textwidth]{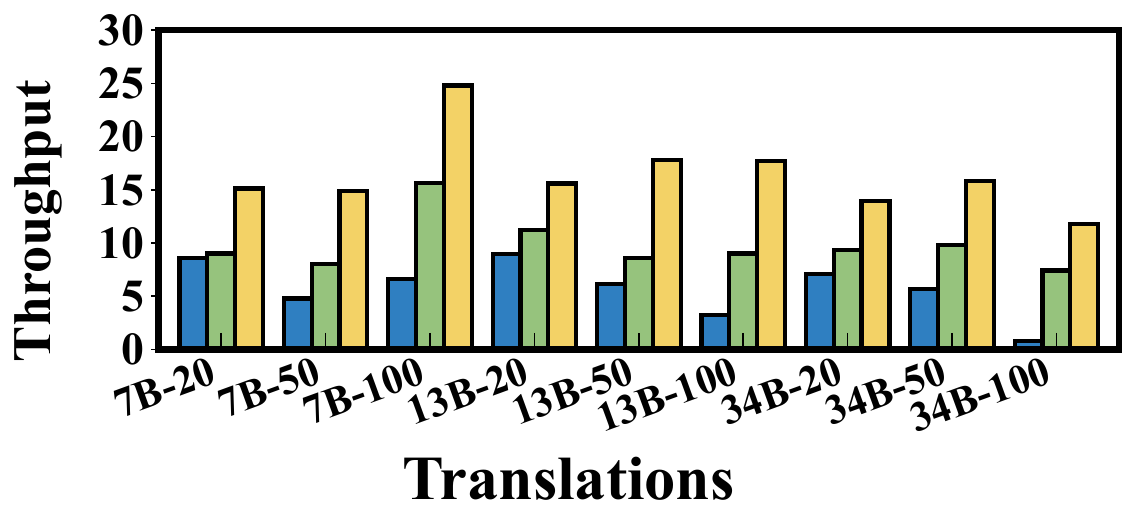}
    \end{subfigure}
    \begin{subfigure}[b]{0.3\textwidth}
        \includegraphics[width=\textwidth]{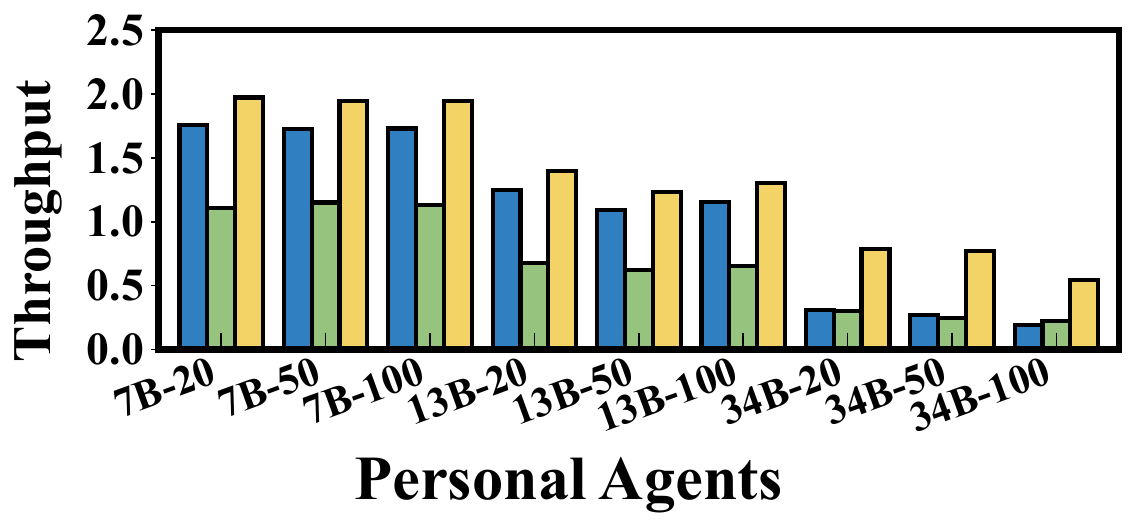}
    \end{subfigure}
    \caption{The average TTFT, TPOT, and supported peak throughput of \sysname{}, vLLM, and S-LoRA in various scenarios. The X-axis represents the model size and LoRA number, e.g., 7B-20 represents Llama-7B with the LoRA number of 20. }
    \label{fig:eval_big} 
    \vspace{-6mm}
\end{figure*}
\section{Evaluation of \sysname{}\label{eval}}
In this section, we first show the serving performance of \sysname{} under various Multi-LoRA application scenarios.
Then, we dive into the reasons behind the performance gains achieved by \sysname{} and the effectiveness of each module. 

\subsection{Evaluation Setup}
\autoref{table:config} has shown our experimental platform. We use Llama-7B/13B/34B as the base model for our evaluations.
Based on the parameter size, we use 1, 2, and 4 NPU cards to deploy the Llama-7B, Llama-13B, and Llama-34B, respectively.
Following previous work~\cite{sheng2024slora,wu2024dlora}, we construct LoRAs based on the model parameters and set the rank of them to 32 and 64 randomly. 
We construct various numbers (i.e., 20,50, and 100) of LoRAs for each model, and the parameters of the LoRAs are randomly generated using a normal distribution. 



We use 
Multi-LoRA inference systems vLLM~\cite{kwon2023efficient} and S-LoRA~\cite{sheng2024slora} as baselines.
vLLM partitions HBM and allocates static HBM space for LoRAs and KV caches, and uses the prefix-caching to reuse history KV caches. It uses the LRU policy to directly discard the KV caches or LoRAs when HBM is full.
We adapt a swap-out policy based on LRU for vLLM to offload its history KV caches and LoRAs in the main memory when the HBM is full.
Moreover, S-LoRA utilizes a unified caching pool for LoRAs and KV caches, but it does not reuse history KV caches and discards them after the query finishes.
It swaps-in the required LoRAs on-demand and swap-out them when no queries use them.

Following prior works~\cite{zhong2024distserve,wu2024dlora}, we utilize the TTFT, TPOT, and peak throughput as the metrics for Multi-LoRA serving.
The peak throughput is determined as the supported maximum number of queries per second when the TTFT is below 500ms.

\subsection{Application Scenarios}
We construct three commonly-used LLM inference application scenarios based on real-world traces.

\textbf{Chatbots.}
In each round of dialogue, chatbots use all the user's previous history to generate new responses. 
Online services often allow users to choose specific dialogue scenarios, such as business analysis~\cite{tsai2016discovering}, and use Multi-LoRA inference to improve efficiency.
We construct queries based on the LMSYS-33k dataset~\cite{zheng2023judging}, which contains 33,000 dialogues with real human preferences.
Each sample includes the model name, the dialogue text, and the timestamp.
Based on the model name, we generate the target LoRA of each query and maintain the original query distribution for different models. 
Moreover, we proportionally scale the LMSYS-33k dataset to achieve different average query sending rates while preserving the original pattern, similar to previous work~\cite{sheng2024slora,wu2024dlora}.

\textbf{Multi-language Translations.} 
This kind of service use Multi-LoRAs to dynamically select and apply the best model to enhance translation results~\cite{zhu2023multilingual}.
We construct queries based on the OPUS-100 dataset~\cite{zhang2020improving}, which contains 55 million sentence pairs across 100 languages.
We make each language translation pair correspond to a specific LoRA, e.g., from French to English. Since the OPUS-100 dataset lacks timestamps, we sample query arrival patterns from the Microsoft Azure function trace (MAFT)~\cite{shahrad2020serverless,zhang2021faster}, following previous work~\cite{wu2024dlora,iliakopoulou2024chameleon}. 
We sort this trace's functions by invoking frequency, select the top-n types of queries, and map them to the $n$ LoRAs to maintain query distribution.


\textbf{Personal Agents.}
LLMs are widely applied in personal agents to provide customized support, such as mobile smart assistants and home assistants, with Multi-LoRA commonly applied for this multi-task serving~\cite{li2024personal,zhao2024lora}.
We utilize the Google Taskmaster~\cite{byrne-etal-2019-taskmaster} to construct queries, which are designed to train and evaluate task-oriented dialogue systems. It contains multi-turn dialogues with complex context management and information exchange, reflecting real-life interactions with assistants. We apply the same sampling method based on MAFT as in the translation scenario.

To adapt different used LoRA numbers ($n$) to the above three scenarios, 
we randomly choose the query patterns from $n$ models, translation pairs, or task scenes in the corresponding dataset and map them to $n$ LoRAs, respectively.


\subsection{Latency and Peak Throughput}


We first evaluate \sysname{} on the inference latency and the peak throughput in the three application scenarios.
For each scenario, the evaluations are conducted under various models and LoRA numbers. 
For each model with a specific LoRA number, we conduct 10 sets of sending rates from 0 to peak throughput of \sysname{}, then we collect the average TTFT and TPOT of them. 
\autoref{fig:eval_big} shows the average TTFT, TPOT, and peak supported throughput of each model with each LoRA number under \sysname{}, vLLM, and S-LoRA.

As observed, \sysname{} reduces the TTFT and TPOT, as well as improves the peak throughput in all the test cases. The average reduction of TTFT and TPOT is 60.3\% and 33.9\% compared to vLLM, and 50.1\% and 28.6\% compared to S-LoRA.
The average peak throughput of \sysname{} is 1.7X and 1.6X of vLLM and S-LoRA, respectively.
The performance increase of \sysname{} originates from maintaining the usage dependencies between LoRAs and KV caches and retaining the most beneficial LoRAs and KV caches in HBM to eliminate the cold-start overhead.
The decrease of TPOT of \sysname{} is smaller than the TTFT compared to baselines because the cold-start overhead mainly impacts more on the prefill stage of LLM inference, which directly leads to increased TTFT. 

Compared to vLLM, \sysname{} decreases more TTFT (average 68.9\%) in the translation scenario than in other scenarios (average 52.5\%). This is because the distribution of LoRAs in this scenario varies more with the OPUS-100 and MAFT datasets.
vLLM's static HBM partition results in poorer cache management, while \sysname{} maintains consistent performance.
Compared to S-LoRA, \sysname{} achieves the best TTFT reduction (average 81.1\%) in the personal agent than others (average 34.6\%). This is because this scenario has the longest average conversation length, and S-LoRA's drawback of not retaining history KVs is signified.
Similarly, while S-LoRA outperforms vLLM in chatbot and translation scenarios due to larger LoRA distribution changes, it struggles in personal agents due to the long conversation.



\begin{figure}
    \centering
    \includegraphics[width=.7\linewidth]{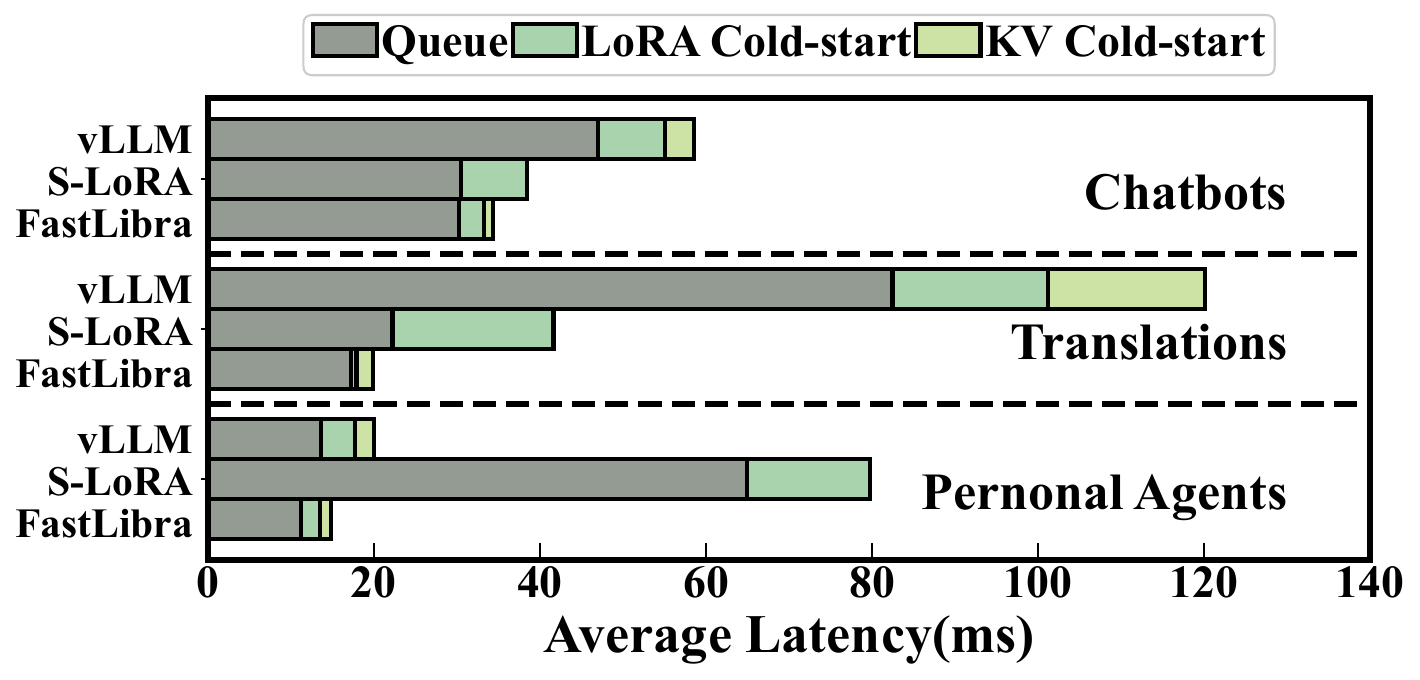}
    \caption{The breakdown of the average queue, LoRA cold-start, and KV cold-start latency in TTFT.}
    \label{fig:eval_ttft_break} 
    \vspace{-3mm}
\end{figure}

\subsection{Diving into the High Serving Performance}
In this section, we show the breakdown of TTFT, the HBM utilization, and the cache hit rate of \sysname{} and baselines, to dive into the reasons for \sysname{}'s high serving performance.




\begin{figure}
    \centering

    \begin{subfigure}{0.48\linewidth} 
        \centering
        \includegraphics[width=\linewidth]{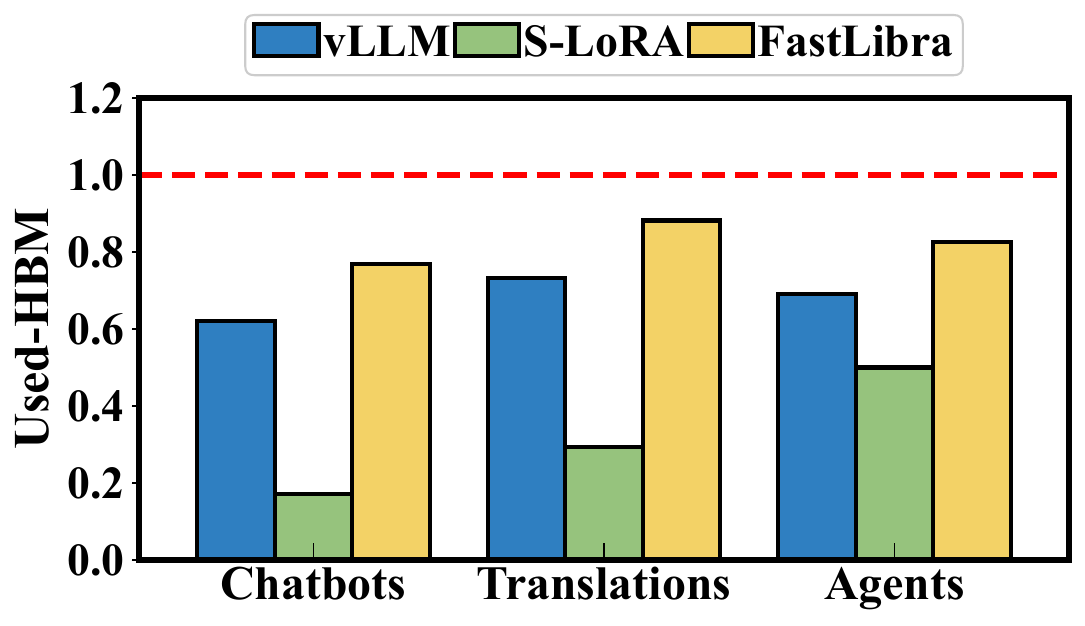}
        \caption{HBM Utilization.}
        \label{fig:hbm_usage}
    \end{subfigure}
    \hfill 
    \begin{subfigure}{0.5\linewidth} 
        \centering
        \includegraphics[width=\linewidth]{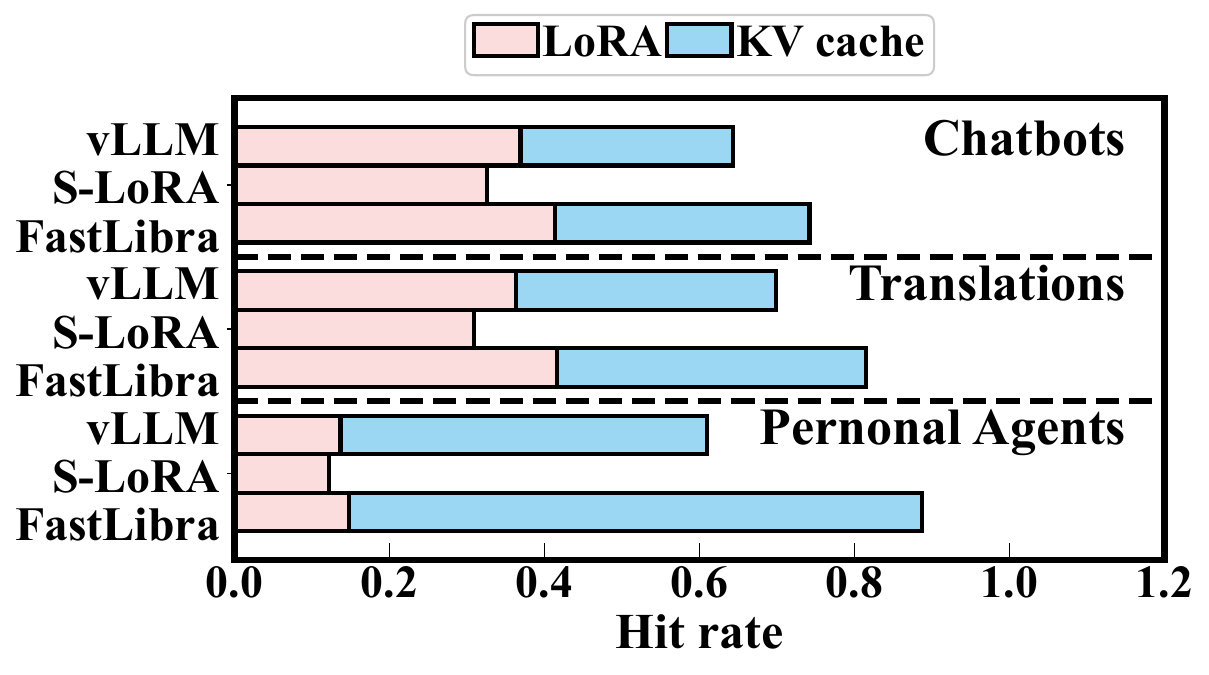}
        \caption{Cache hit rate.}
        \label{fig:hit_rate}
        \vspace{-1mm}
    \end{subfigure}
    
    \caption{The average HBM usage and cache hit rate of \sysname{} and baselines of different scenarios.}
    \label{fig:eval_hbm} 
    \vspace{-3mm}
\end{figure}

\autoref{fig:eval_ttft_break} shows the breakdown of the average queue, LoRA cold-start, and KV cold-start latency in TTFT in different scenarios. 
We can observe that \sysname{} achieves the lowest queue, LoRA cold-start, and KV cold-start latency in all scenarios. 
This means that \sysname{} has the highest HBM utilization efficiency.

For in-depth analysis, we sample the average HBM utilization of \sysname{} and baselines across different scenarios, shown in \autoref{fig:hbm_usage}.
\sysname{} improves HBM utilization by 1.2X and 2.6X over vLLM and S-LoRA, respectively, due to its dynamic swapping of LoRAs and KV caches in a unified caching pool. 
In contrast, S-LoRA wastes HBM by not retaining history KV caches, while vLLM's static HBM partition makes the HBM for LoRAs or KVs under-utilized under dynamic loads.
These factors also contribute to lower queue and cold-start latency for \sysname{}, as shown in \autoref{fig:eval_ttft_break}.





We also compare the average KV cache and LoRA hit rates of \sysname{} and baselines across different scenarios, as shown in \autoref{fig:hit_rate}.
\sysname{} increases the cache hit rate by 1.3X and 3.2X compared to vLLM and S-LoRA, respectively.
This is because \sysname{} maintains the usage dependencies between LoRAs and KV caches to eliminate invalid KV caches which enhances the HBM utilization efficiency. Its efficient swapping strategy also prefetches appropriate KV caches and LoRAs into HBM.
S-LoRA has the lowest hit rate because it does not reuse history KV caches.
As a result, \sysname{} achieves lower queue and cold-start latency for both LoRA and KV caches in \autoref{fig:eval_ttft_break}.

\begin{figure}
    \centering
    \begin{subfigure}{\linewidth}
        \centering
        \includegraphics[width=.85\linewidth]{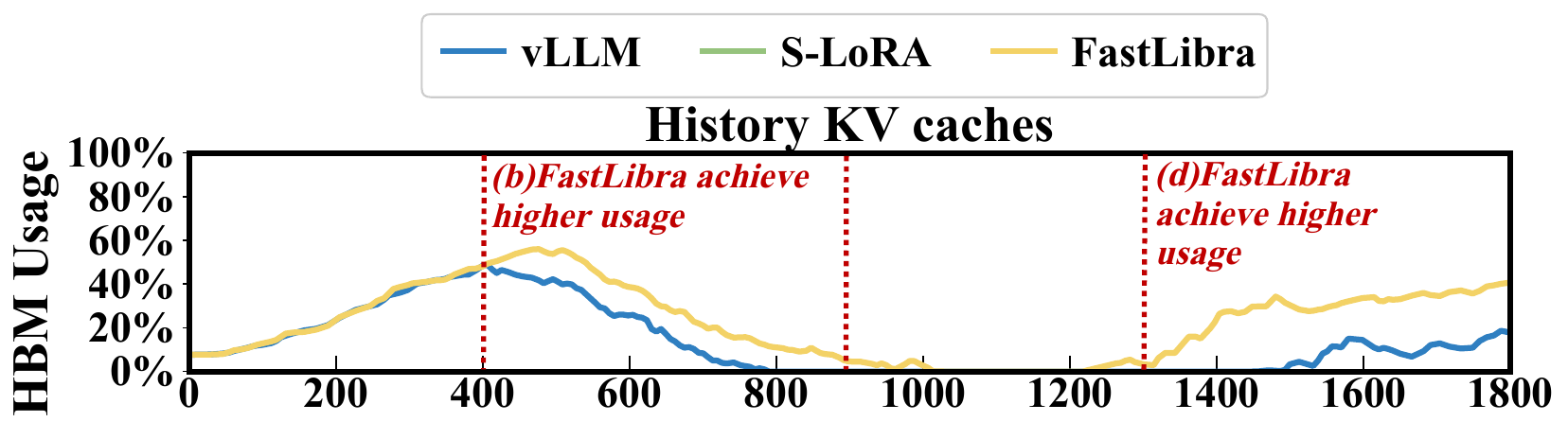}
        \label{fig:dynamic_hbm_ours_best}
    \end{subfigure}
    \vspace{-0mm}
    \begin{subfigure}{\linewidth}
        \centering
        \includegraphics[width=.85\linewidth]{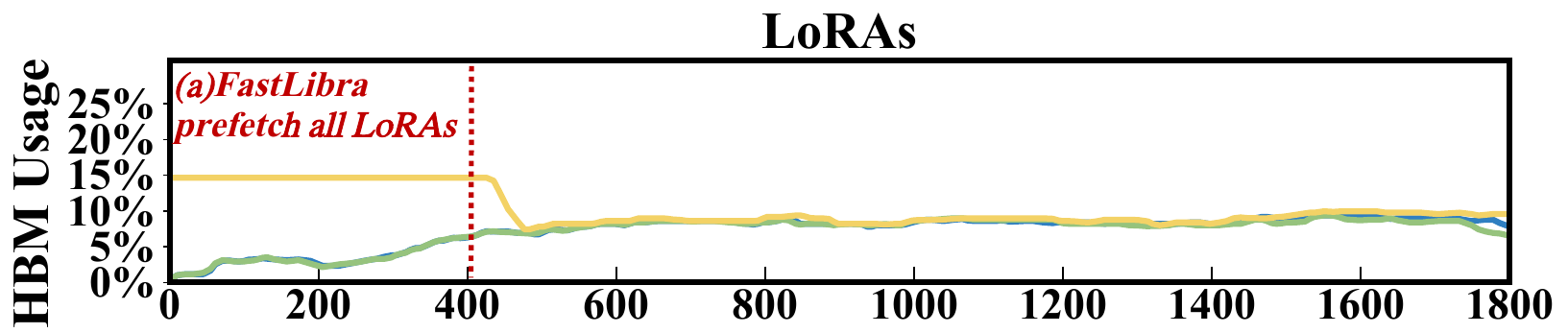}
        \label{fig:dynamic_hbm_vllm}
    \end{subfigure}
    \vspace{-0mm}
    \begin{subfigure}{\linewidth}
        \centering
        \includegraphics[width=.85\linewidth]{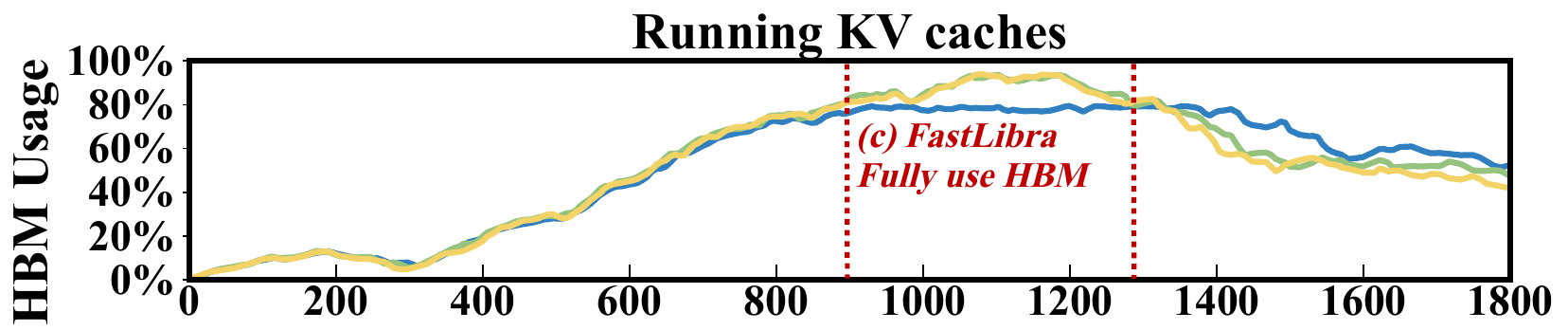}
        \label{fig:dynamic_hbm_slora}
    \end{subfigure}
    \vspace{-0mm}
    \begin{subfigure}{\linewidth}
        \centering
        \includegraphics[width=.85\linewidth]{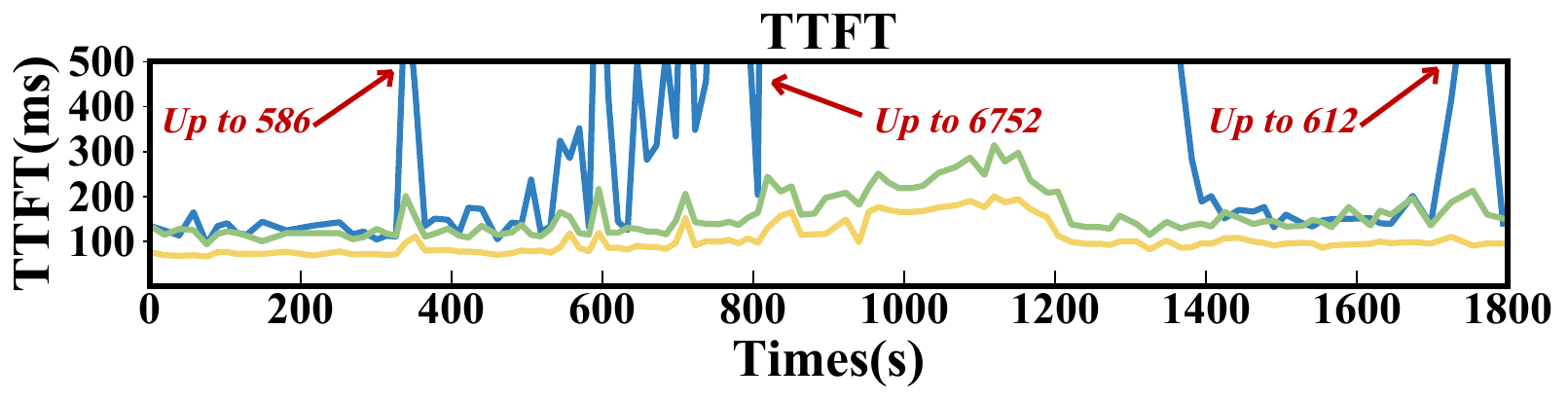}
        \label{fig:dynamic_ttft}
    \end{subfigure}
    
    \caption{The HBM allocation over time of \sysname{} and baselines in different application scenarios.}
    \label{fig:dynamic_hbm} 
    \vspace{-3mm}
\end{figure}

\begin{figure*}
    \centering

    \begin{subfigure}{0.33\linewidth} 
        \centering
        \includegraphics[width=\linewidth]{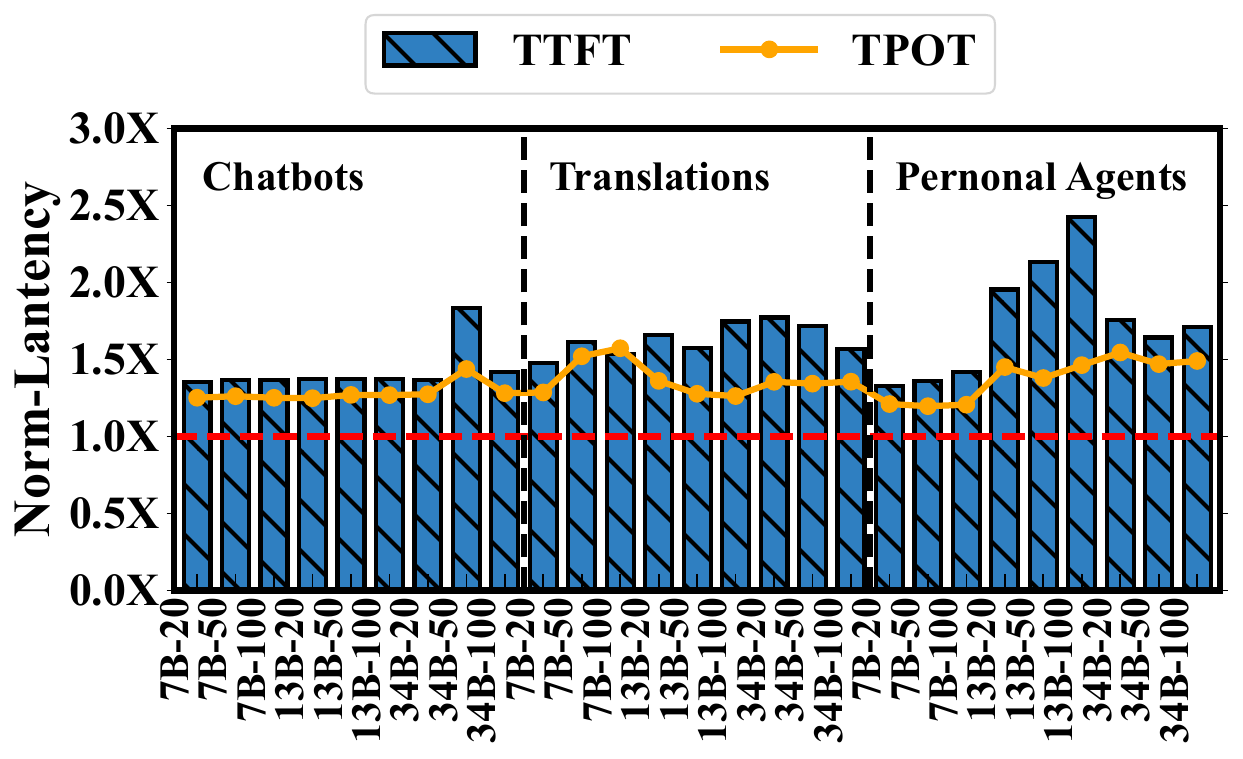}
        \caption{\sysname{}-WOM}
        \label{fig:wom}
    \end{subfigure}
    \hfill 
    \begin{subfigure}{0.33\linewidth} 
        \centering
        \includegraphics[width=\linewidth]{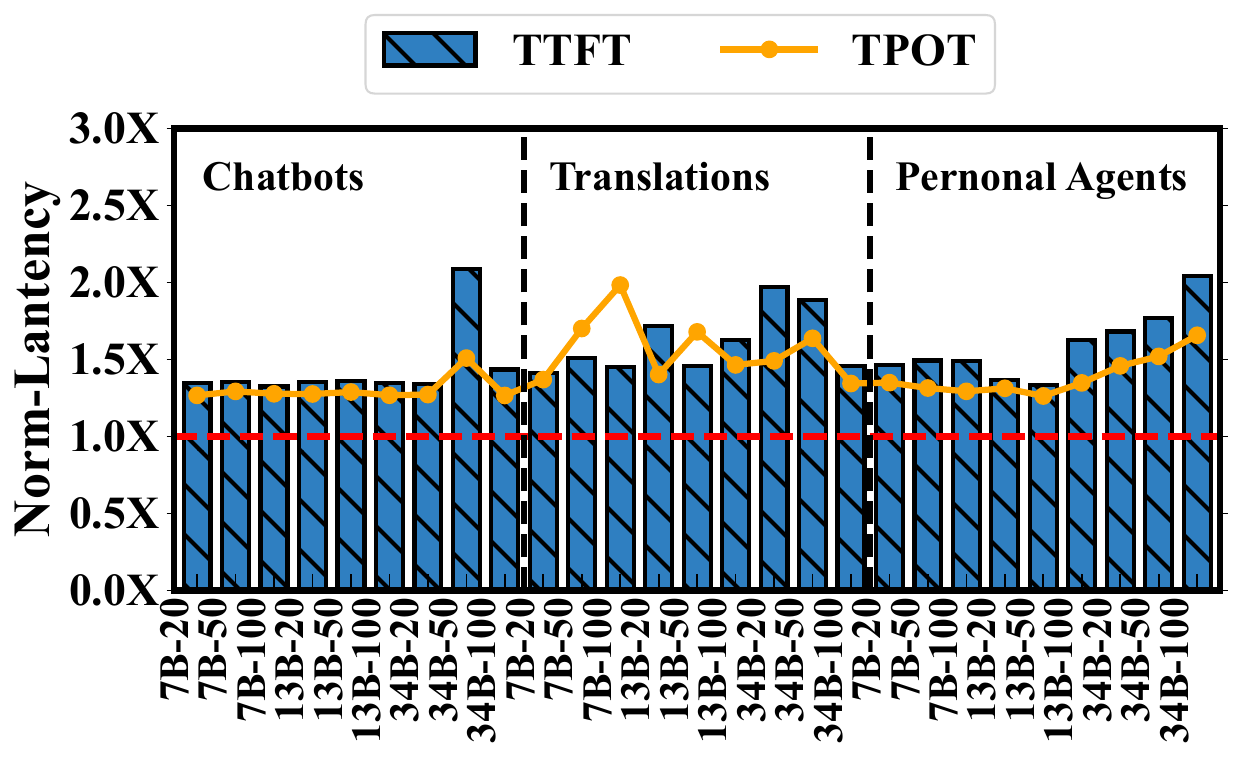}
        \caption{\sysname{}-WOS}
        \label{fig:wos}
    \end{subfigure}
    \hfill 
    \begin{subfigure}{0.33\linewidth} 
        \centering
        \includegraphics[width=\linewidth]{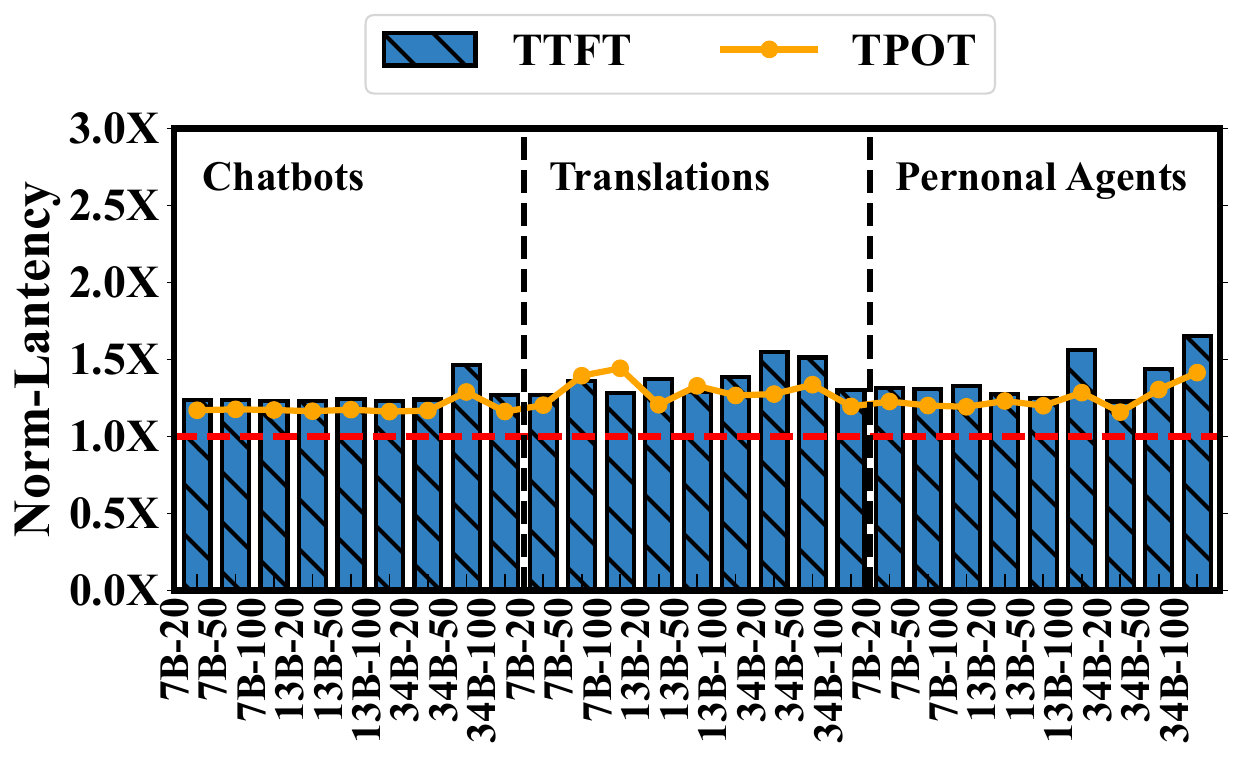}
        \caption{\sysname{}-WOL}
        \label{fig:wol}
    \end{subfigure}
    
    \caption{The TTFT and TPOT of the variants of \sysname{} in different application scenarios, respectively.}
    \label{fig:abl} 
     \vspace{-6mm}
\end{figure*}

\subsection{Investigating HBM Allocation Over Time\label{sec:hbm_breakdown}}
In this subsection, we compare HBM allocation between \sysname{} and baselines to show the effectiveness of \sysname{}'s cache management. We take the example of using Llama-13B model, LoAR number of 100, and average sending rate of 1.6 for the chatbot scenario. Other scenarios have similar results. \autoref{fig:dynamic_hbm} shows the HBM allocations for history KV caches, LoRAs, and running KV caches under \sysname{} and baselines.


From 0s to 400s shown in \textbf{(a)}, \sysname{} proactively fetches all LoRAs into HBM based on the cost model to eliminate the cold-start overhead of LoRAs under low HBM pressure. In contrast, vLLM and S-LoRA load LoRAs on-demand, leading to higher TTFT in this period.
From 400s to 900s shown in \textbf{(b)}, as the query sending rate increases, \sysname{} swaps out some LoRAs and retains the most history KV caches in HBM due to the unified caching pool. 
In contrast, vLLM's static HBM partition retains fewer history KVs while S-LoRA directly discards them, leading to poorer KV cache reuse and higher TTFT.
Moreover, history KVs gradually decrease in this period as they are swapped out to free HBM for running KVs when the sending rate rises.

From 900s to 1300s in \textbf{(c)}, \sysname{} swaps-out all history KV caches to free up HBM for running KV caches of the current inference.
By contrast, the static HBM partition of vLLM results in the KV cache memory pool being exhausted to its maximum capacity (80\%), leading to queuing and the rapid growth of TTFT.
At last, from 1300s to 1800s in \textbf{(d)}, \sysname{} can retain more history KV caches than vLLM and S-LoRA, leading to a higher HBM usage. During this period, vLLM has higher running KV caches than S-LoRA and \sysname{} because of the previous query queuing.

\subsection{Effectiveness of the Cache Manager}
In this subsection, we show the performance of \sysname{}-WOM, a variant of \sysname{} that does not maintain usage dependencies between LoRAs and KV caches with the cache manager. 
The \sysname{}-WOM still uses the cache swapper to swap-in/out LoRAs or KVs in the unified caching pool.

\autoref{fig:wom} shows the TTFT and TPOT of \sysname{}-WOM normalized to \sysname{}. As observed, the TTFT and TPOT of \sysname{}-WOM are higher than \sysname{} in all cases, with an average increase of 1.27X and 1.18X, respectively.
We also sample the history KV caches during the inference and find \sysname{}-WOM suffers from an average of 48.6\% invalid KV caches.
Moreover, the peak supported throughput of \sysname{}-WOM is decreased by 19.8\% compared to \sysname{}.

When ignoring the usage dependencies between LoRAs and KV caches, lots of invalid KV caches occupy the HBM space but cannot be matched due to their front LoRAs are not loaded, leading to low HBM utilization efficiency. In this case, the useful LoRAs or KV caches cannot be loaded, thus leading to low query serving performance.






\subsection{Effectiveness of the Cache Swapper}
In this subsection, we show the performance of \sysname{}-WOS, a variant of \sysname{} that uses a simple LRU policy to replace the cost model (\autoref{eq:C_swap}) in the cache swapper. 
The usage dependencies between LoRAs and KV caches are still maintained with the cache manager during inference.

\autoref{fig:wos} shows the TTFT and TPOT of \sysname{}-WOS normalized to \sysname{}. We can observe that both the TTFT and TPOT of \sysname{}-WOS are increased in all test cases, with an average increase of 1.24X and 1.15X, respectively.
Moreover, the supported peak throughput of \sysname{}-WOS is also decreased by 17.2\%.

Without \sysname{}'s cost model to access the benefits or harms to TTFT for swap-in/out, inappropriate LoRAs or KV caches will be swapped-in/out when HBM is idle/busy. This results in more cold-start overheads of LoRAs and KVs, thus decreasing the serving performance. 




\subsection{Effectiveness of the Enough LoRAs}
In this subsection, we show the performance of \sysname{}-WOL, a variant of \sysname{} that ignores the required LoRA quantity of the cache swapper. Thus, when evaluating the benefits of retaining a node in HBM, \sysname{}-WOL eliminates the LoRA reward (\autoref{eq:lora_demand}) in the cost model.

\autoref{fig:wol} shows the TTFT and TPOT of \sysname{}-WOL normalized to \sysname{}. 
\sysname{}-WOL increases the TTFT and TPOT in all the test cases, with an average increase of 1.13X and 1.11X, respectively. With the \sysname{}-WOL, the peak supported throughput is also decreased by 13.1\% on average.
Compared to \sysname{}-WOS, \sysname{}-WOL's serving performance is increased as it considers part of our cost model (\autoref{eq:reward_swapin}), but still has a gap to the \sysname{}.

The insufficient LoRA loading in some dynamic scenarios can lead to a large number of LoRA cold-starts. As each query inference can only start once the required LoRA is matched in HBM, it can leads to the increase of TTFT.

\begin{figure}
    \centering
    \includegraphics[width=.9\linewidth]{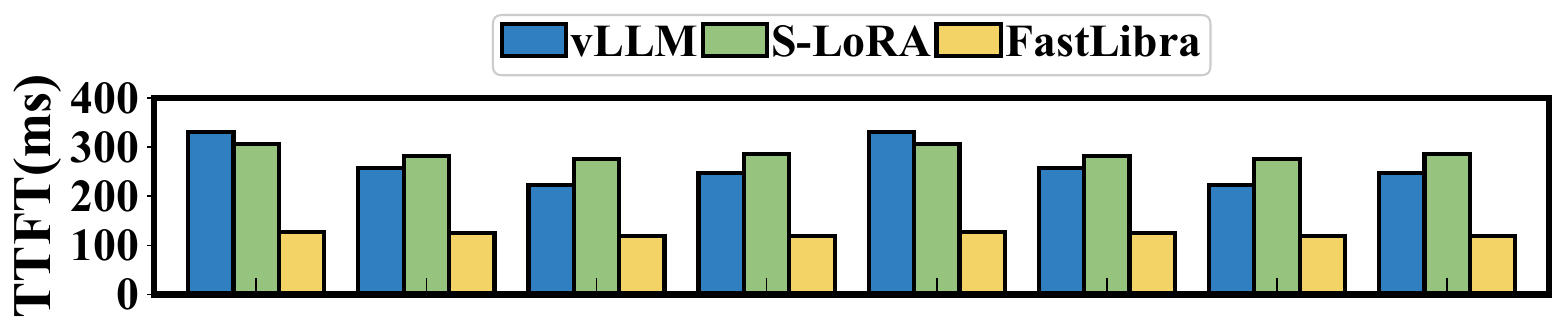}
    \includegraphics[width=.9\linewidth]{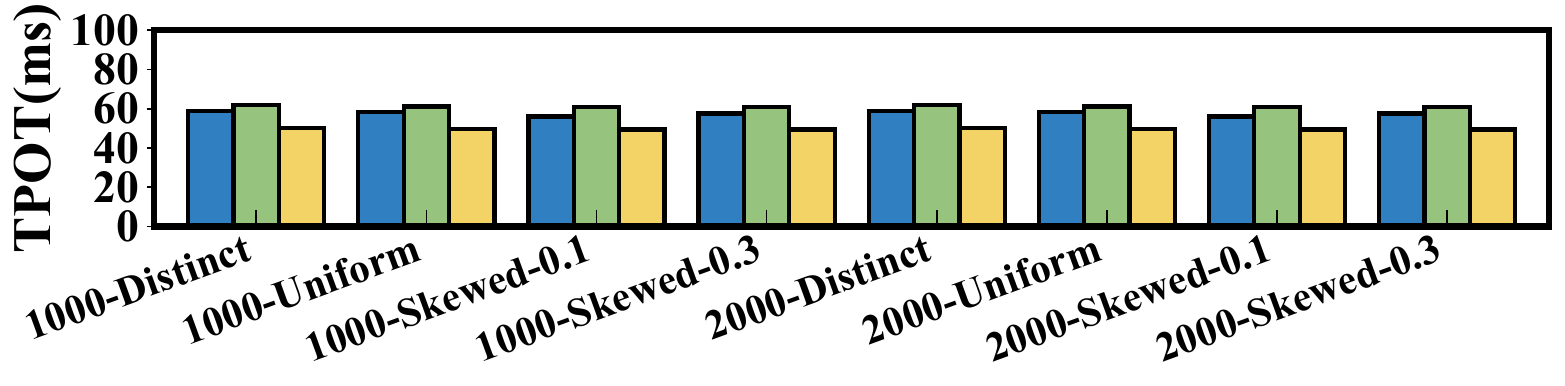}
    \caption{The TTFT and TPOT of \sysname{} and baselines with different LoRA numbers and distributions. The x-axis represents the combination of LoRA number and distribution.}
    \label{fig:extend} 
    \vspace{-3mm}
\end{figure}

\subsection{The Impacts of a Large Number of LoRAs}
In this subsection, we investigate the effectiveness of \sysname{} when thousands of LoRAs exist, although real-world scenarios always only have tens of LoRAs~\cite{zhao2024lora,apple-assistant}.
we use the Llama-7B under the chatbot scenario as an example. The LoRA number is 1000 or 2000, and we set three types of LoRA distributions: 1) \textit{Uniform}, where queries have an equal usage probability for each LoRA. 2) \textit{Distinct}, where queries are handled by polling to use a LoRA. 3) \textit{Skewed-$x$}, where we construct queries using different LoRAs based on Gaussian distribution and set different standard deviations $x$.

\autoref{fig:extend} shows the TTFT and TPOT of \sysname{}, vLLM, and S-LoRA, respectively.
We can observe that \sysname{} has both the lower TTFT and TPOT in all the test cases, with an average decrease of 55.4\% and 16.2\%, respectively.
Specially, vLLM's and S-LoRA's TTFT varies obviously across different scenarios, while \sysname{} can always maintain a stable and low level of TTFT.
The above results prove the generality of \sysname{} under the large number of LoRAs with different distributions.



\subsection{Overhead of \sysname{}}
The overheads of \sysname{} mainly come from three parts: the dependency tree matching and updating in the cache manager, the monitoring of the HBM usage, and the swapping decisions of the cache swapper.
For the cache manager, we employ an efficient trie tree for rapid matching and updating. Even if the HBM resources are fully utilized and the size of tree reaches the maximum, the average overhead for matching and updating is less than 0.5ms. 
Moreover, for monitoring HBM usage and the swapping decisions of the cache swapper, the time overhead for them can be done within 5ms.

The above overheads are all acceptable relative to the entire inference process of each query, which can take seconds or even tens of seconds.


\section{Related Work}

\textbf{LLM Fine-tuning.}
Recent studies have proposed efficient methods for fine-tuning large language models~\cite{hu2021lora,li2021prefix,lester2021power}, with LoRA adapters being among the most widely used.
LoRA achieves fine-tuning with low costs by adding a low-rank branch~\cite{hu2021lora}.
Moreover, evolved models like DoRA~\cite{liu2024dora} and AdaLoRA~\cite{zhang2023adalora} that developed based on LoRA, enhance fine-tuning by introducing flexible updates through weight decomposition and efficient trimming of insignificant singular values.
While improving fine-tuning efficiency, these models share the same features as native LoRA for query inference, i.e., adding branches to the original transformer layers with similar computation and memory patterns, thus \sysname{} can adapt to them with minimal modifications.

\textbf{KV Cache Management.}
Original LLM inference engines like Orca~\cite{yu2022orca} and FastTransformer~\cite{shazeer2019fast} directly discard requested KV caches after query processing.
To reduce the recomputations, SGLang~\cite{zheng2023efficiently} introduced RadixAttention, which reuses history KV caches using a global prefix tree and LRU strategy.
ChunkAttention~\cite{zheng2023efficiently} further improves HBM utilization by sharing KV caches for common prefixes across queries at runtime.
For multi-round conversations, AttentionStore~\cite{gao2024attentionstore} and Pensieve~\cite{yu2023stateful} maintained a multi-level KV cache system to store and manage all requested history KV caches to eliminate recalculations.
Although the above works can reuse history KV caches to improve query inference performance, they neglect to unified manage KV caches along with LoRAs in the HBM under the Multi-LoRA scenario.

\textbf{Multi-LoRA Serving.}
Several inference systems have been proposed for Multi-LoRA serving. 
S-LoRA~\cite{sheng2024slora} and Punica~\cite{chen2024punica} separated the base model from the task-specific adapter and dynamically loaded them into HBM. They utilized customized operators to realize that queries using different LoRAs can be batched to improve inference efficiency.
dLoRA~\cite{wu2024dlora} dynamically switched between merged and unmerged modes to reduce the inference latency, which is orthogonal to the work in this paper. 
These works did not consider caching history KV caches to avoid recomputation. 
Moreover, vLLM~\cite{vLLM2024code} and SGLang~\cite{sglang2024code} integrated S-LoRA's operators for batched Multi-LoRA inference with static HBM allocation and LRU eviction for LoRAs and KV caches. 
However, they managed LoRAs and KV caches separately, failing to account for their usage dependencies and balance the HBM usage, resulting in poor Multi-LoRA serving performance.







%

\section{Conclusion}
In this paper, we propose \sysname{} to optimize the caching of LoRAs and KV caches to improve the Multi-LoRA serving performance.
\sysname{}'s cache manager maintains the usage dependencies between KV caches and LoRAs based on a tree-based scheme with a unified caching pool. Based on this scheme, the invalid KV caches in the HBM can be eliminated to improve the HBM utilization efficiency.
\sysname{}'s cache swapper periodically determines the swap-in/out of LoRAs and KVs by using a unified cost model which reflects the benefits of swap-in/out LoRA and KVs to the performance of future queries.
We have implemented \sysname{} on top of vLLM and experimental results show \sysname{} reduces the TTFT and TPOT by 63.4\% and 40.1\% on average, respectively, compared to state-of-the-art works.


\bibliographystyle{plain}
\bibliography{references}

\begin{thebibliography}{10}

\bibitem{agrawal2023sarathi}
Amey Agrawal, Ashish Panwar, Jayashree Mohan, Nipun Kwatra, Bhargav~S Gulavani, and Ramachandran Ramjee.
\newblock Sarathi: Efficient llm inference by piggybacking decodes with chunked prefills.
\newblock {\em arXiv preprint arXiv:2308.16369}, 2023.

\bibitem{alizadeh2023llm}
Keivan Alizadeh, Iman Mirzadeh, Dmitry Belenko, Karen Khatamifard, Minsik Cho, Carlo~C Del~Mundo, Mohammad Rastegari, and Mehrdad Farajtabar.
\newblock Llm in a flash: Efficient large language model inference with limited memory.
\newblock {\em arXiv preprint arXiv:2312.11514}, 2023.

\bibitem{apple-assistant}
Apple.
\newblock Introducing apple’s on-device and server foundation models, 2025.

\bibitem{aryan2023costly}
Abi Aryan, Aakash~Kumar Nain, Andrew McMahon, Lucas~Augusto Meyer, and Harpreet~Singh Sahota.
\newblock The costly dilemma: generalization, evaluation and cost-optimal deployment of large language models.
\newblock {\em arXiv preprint arXiv:2308.08061}, 2023.

\bibitem{byrne-etal-2019-taskmaster}
Bill Byrne, Karthik Krishnamoorthi, Chinnadhurai Sankar, Arvind Neelakantan, Daniel Duckworth, Semih Yavuz, Ben Goodrich, Amit Dubey, Kyu-Young Kim, and Andy Cedilnik.
\newblock Taskmaster-1:toward a realistic and diverse dialog dataset.
\newblock In {\em 2019 Conference on Empirical Methods in Natural Language Processing and 9th International Joint Conference on Natural Language Processing}, Hong Kong, 2019.

\bibitem{chen2024punica}
Lequn Chen, Zihao Ye, Yongji Wu, Danyang Zhuo, Luis Ceze, and Arvind Krishnamurthy.
\newblock Punica: Multi-tenant lora serving.
\newblock {\em Proceedings of Machine Learning and Systems}, 6:1--13, 2024.

\bibitem{chiang2024chatbot}
Wei-Lin Chiang, Lianmin Zheng, Ying Sheng, Anastasios~Nikolas Angelopoulos, Tianle Li, Dacheng Li, Hao Zhang, Banghua Zhu, Michael Jordan, Joseph~E Gonzalez, et~al.
\newblock Chatbot arena: An open platform for evaluating llms by human preference.
\newblock {\em arXiv preprint arXiv:2403.04132}, 2024.

\bibitem{chowdhery2023palm}
Aakanksha Chowdhery, Sharan Narang, Jacob Devlin, Maarten Bosma, Gaurav Mishra, Adam Roberts, Paul Barham, Hyung~Won Chung, Charles Sutton, Sebastian Gehrmann, et~al.
\newblock Palm: Scaling language modeling with pathways.
\newblock {\em Journal of Machine Learning Research}, 24(240):1--113, 2023.

\bibitem{google-tpu-intro}
Google Cloud.
\newblock Introduction to tpus, 2023.

\bibitem{sglang2024code}
SGL Community.
\newblock sglang: A fast serving framework for large language models and vision language models., 2024.

\bibitem{hbm-wikipedia}
Wikipedia contributors.
\newblock High bandwidth memory, 2023.

\bibitem{trie-wikipedia}
Wikipedia contributors.
\newblock Trie, 2023.

\bibitem{nvidia-a100}
NVIDIA Corporation.
\newblock Nvidia a100 tensor core gpu, 2024.

\bibitem{dettmers2024qlora}
Tim Dettmers, Artidoro Pagnoni, Ari Holtzman, and Luke Zettlemoyer.
\newblock Qlora: Efficient finetuning of quantized llms.
\newblock {\em Advances in Neural Information Processing Systems}, 36, 2024.

\bibitem{floridi2020gpt}
Luciano Floridi and Massimo Chiriatti.
\newblock Gpt-3: Its nature, scope, limits, and consequences.
\newblock {\em Minds and Machines}, 30:681--694, 2020.

\bibitem{gao2024attentionstore}
Bin Gao, Zhuomin He, Puru Sharma, Qingxuan Kang, Djordje Jevdjic, Junbo Deng, Xingkun Yang, Zhou Yu, and Pengfei Zuo.
\newblock Attentionstore: Cost-effective attention reuse across multi-turn conversations in large language model serving.
\newblock {\em arXiv preprint arXiv:2403.19708}, 2024.

\bibitem{gim2024prompt}
In~Gim, Guojun Chen, Seung-seob Lee, Nikhil Sarda, Anurag Khandelwal, and Lin Zhong.
\newblock Prompt cache: Modular attention reuse for low-latency inference.
\newblock {\em Proceedings of Machine Learning and Systems}, 6:325--338, 2024.

\bibitem{Bard}
Google.
\newblock Bard, 2023.

\bibitem{hu2021lora}
Edward~J Hu, Yelong Shen, Phillip Wallis, Zeyuan Allen-Zhu, Yuanzhi Li, Shean Wang, Lu~Wang, and Weizhu Chen.
\newblock Lora: Low-rank adaptation of large language models.
\newblock {\em arXiv preprint arXiv:2106.09685}, 2021.

\bibitem{huang2023lorahub}
Chengsong Huang, Qian Liu, Bill~Yuchen Lin, Tianyu Pang, Chao Du, and Min Lin.
\newblock Lorahub: Efficient cross-task generalization via dynamic lora composition.
\newblock {\em arXiv preprint arXiv:2307.13269}, 2023.

\bibitem{iliakopoulou2024chameleon}
Nikoleta Iliakopoulou, Jovan Stojkovic, Chloe Alverti, Tianyin Xu, Hubertus Franke, and Josep Torrellas.
\newblock Chameleon: Adaptive caching and scheduling for many-adapter llm inference environments.
\newblock {\em arXiv preprint arXiv:2411.17741}, 2024.

\bibitem{kwon2023efficient}
Woosuk Kwon, Zhuohan Li, Siyuan Zhuang, Ying Sheng, Lianmin Zheng, Cody~Hao Yu, Joseph Gonzalez, Hao Zhang, and Ion Stoica.
\newblock Efficient memory management for large language model serving with pagedattention.
\newblock In {\em Proceedings of the 29th Symposium on Operating Systems Principles}, pages 611--626, 2023.

\bibitem{lester2021power}
Brian Lester, Rami Al-Rfou, and Noah Constant.
\newblock The power of scale for parameter-efficient prompt tuning.
\newblock {\em arXiv preprint arXiv:2104.08691}, 2021.

\bibitem{li2024caraserve}
Suyi Li, Hanfeng Lu, Tianyuan Wu, Minchen Yu, Qizhen Weng, Xusheng Chen, Yizhou Shan, Binhang Yuan, and Wei Wang.
\newblock Caraserve: Cpu-assisted and rank-aware lora serving for generative llm inference.
\newblock {\em arXiv preprint arXiv:2401.11240}, 2024.

\bibitem{li2021prefix}
Xiang~Lisa Li and Percy Liang.
\newblock Prefix-tuning: Optimizing continuous prompts for generation.
\newblock {\em arXiv preprint arXiv:2101.00190}, 2021.

\bibitem{li2024personal}
Yuanchun Li, Hao Wen, Weijun Wang, Xiangyu Li, Yizhen Yuan, Guohong Liu, Jiacheng Liu, Wenxing Xu, Xiang Wang, Yi~Sun, et~al.
\newblock Personal llm agents: Insights and survey about the capability, efficiency and security.
\newblock {\em arXiv preprint arXiv:2401.05459}, 2024.

\bibitem{liu2024dora}
Shih-Yang Liu, Chien-Yi Wang, Hongxu Yin, Pavlo Molchanov, Yu-Chiang~Frank Wang, Kwang-Ting Cheng, and Min-Hung Chen.
\newblock Dora: Weight-decomposed low-rank adaptation.
\newblock {\em arXiv preprint arXiv:2402.09353}, 2024.

\bibitem{alpacalora}
Alpaca lora team.
\newblock Instruct-tune llama on consumer hardware using alpaca-lora, 2023.

\bibitem{luo2024arena}
Haipeng Luo, Qingfeng Sun, Can Xu, Pu~Zhao, Qingwei Lin, Jianguang Lou, Shifeng Chen, Yansong Tang, and Weizhu Chen.
\newblock Arena learning: Build data flywheel for llms post-training via simulated chatbot arena.
\newblock {\em arXiv preprint arXiv:2407.10627}, 2024.

\bibitem{mann2020language}
Ben Mann, N~Ryder, M~Subbiah, J~Kaplan, P~Dhariwal, A~Neelakantan, P~Shyam, G~Sastry, A~Askell, S~Agarwal, et~al.
\newblock Language models are few-shot learners.
\newblock {\em arXiv preprint arXiv:2005.14165}, 1, 2020.

\bibitem{martinez2024impact}
Matias Martinez.
\newblock The impact of hyperparameters on large language model inference performance: An evaluation of vllm and huggingface pipelines.
\newblock {\em arXiv preprint arXiv:2408.01050}, 2024.

\bibitem{ChatGPT}
OpenAI.
\newblock Chatgpt, 2020.

\bibitem{torch-stream-docs}
{PyTorch Contributors}.
\newblock torch.stream --- pytorch 2.0.1 documentation, 2023.

\bibitem{qin2024mooncake}
Ruoyu Qin, Zheming Li, Weiran He, Mingxing Zhang, Yongwei Wu, Weimin Zheng, and Xinran Xu.
\newblock Mooncake: Kimi's kvcache-centric architecture for llm serving.
\newblock {\em arXiv e-prints}, pages arXiv--2407, 2024.

\bibitem{shahrad2020serverless}
Mohammad Shahrad, Rodrigo Fonseca, Inigo Goiri, Gohar Chaudhry, Paul Batum, Jason Cooke, Eduardo Laureano, Colby Tresness, Mark Russinovich, and Ricardo Bianchini.
\newblock Serverless in the wild: Characterizing and optimizing the serverless workload at a large cloud provider.
\newblock In {\em 2020 USENIX annual technical conference (USENIX ATC 20)}, pages 205--218, 2020.

\bibitem{shazeer2019fast}
Noam Shazeer.
\newblock Fast transformer decoding: One write-head is all you need.
\newblock {\em arXiv preprint arXiv:1911.02150}, 2019.

\bibitem{sheng2024slora}
Ying Sheng, Shiyi Cao, Dacheng Li, Coleman Hooper, Nicholas Lee, Shuo Yang, Christopher Chou, Banghua Zhu, Lianmin Zheng, Kurt Keutzer, et~al.
\newblock Slora: Scalable serving of thousands of lora adapters.
\newblock {\em Proceedings of Machine Learning and Systems}, 6:296--311, 2024.

\bibitem{staudemeyer2019understanding}
Ralf~C Staudemeyer and Eric~Rothstein Morris.
\newblock Understanding lstm--a tutorial into long short-term memory recurrent neural networks.
\newblock {\em arXiv preprint arXiv:1909.09586}, 2019.

\bibitem{touvron2023llama}
Hugo Touvron, Louis Martin, Kevin Stone, Peter Albert, Amjad Almahairi, Yasmine Babaei, Nikolay Bashlykov, Soumya Batra, Prajjwal Bhargava, Shruti Bhosale, et~al.
\newblock Llama 2: Open foundation and fine-tuned chat models.
\newblock {\em arXiv preprint arXiv:2307.09288}, 2023.

\bibitem{tsai2016discovering}
Ming-Feng Tsai, Chuan-Ju Wang, and Po-Chuan Chien.
\newblock Discovering finance keywords via continuous-space language models.
\newblock {\em ACM Transactions on Management Information Systems (TMIS)}, 7(3):1--17, 2016.

\bibitem{vLLM2024code}
vLLM Community.
\newblock vllm: A high-throughput and memory-efficient inference and serving engine for llms.

\bibitem{wang2024lora}
Zhengbo Wang, Jian Liang, Ran He, Zilei Wang, and Tieniu Tan.
\newblock Lora-pro: Are low-rank adapters properly optimized?
\newblock {\em arXiv preprint arXiv:2407.18242}, 2024.

\bibitem{wu2024dlora}
Bingyang Wu, Ruidong Zhu, Zili Zhang, Peng Sun, Xuanzhe Liu, and Xin Jin.
\newblock $\{$dLoRA$\}$: Dynamically orchestrating requests and adapters for $\{$LoRA$\}$$\{$LLM$\}$ serving.
\newblock In {\em 18th USENIX Symposium on Operating Systems Design and Implementation (OSDI 24)}, pages 911--927, 2024.

\bibitem{yao2024cacheblend}
Jiayi Yao, Hanchen Li, Yuhan Liu, Siddhant Ray, Yihua Cheng, Qizheng Zhang, Kuntai Du, Shan Lu, and Junchen Jiang.
\newblock Cacheblend: Fast large language model serving with cached knowledge fusion.
\newblock {\em arXiv preprint arXiv:2405.16444}, 2024.

\bibitem{yu2022orca}
Gyeong-In Yu, Joo~Seong Jeong, Geon-Woo Kim, Soojeong Kim, and Byung-Gon Chun.
\newblock Orca: A distributed serving system for transformer-based generative models.
\newblock In {\em 16th USENIX Symposium on Operating Systems Design and Implementation (OSDI 22)}, pages 521--538, 2022.

\bibitem{yu2023stateful}
Lingfan Yu, Jinkun Lin, and Jinyang Li.
\newblock Stateful large language model serving with pensieve.
\newblock {\em arXiv preprint arXiv:2312.05516}, 2023.

\bibitem{zhang2020improving}
Biao Zhang, Philip Williams, Ivan Titov, and Rico Sennrich.
\newblock Improving massively multilingual neural machine translation and zero-shot translation.
\newblock {\em arXiv preprint arXiv:2004.11867}, 2020.

\bibitem{zhang2016google}
Bill Zhang.
\newblock Google’s neural machine translation system: Bridging the gap between human and machine translation.
\newblock {\em arXiv preprint arXiv:1609.08144}, 11, 2016.

\bibitem{zhang2023adalora}
Qingru Zhang, Minshuo Chen, Alexander Bukharin, Nikos Karampatziakis, Pengcheng He, Yu~Cheng, Weizhu Chen, and Tuo Zhao.
\newblock Adalora: Adaptive budget allocation for parameter-efficient fine-tuning.
\newblock {\em arXiv preprint arXiv:2303.10512}, 2023.

\bibitem{zhang2021faster}
Yanqi Zhang, {\'I}{\~n}igo Goiri, Gohar~Irfan Chaudhry, Rodrigo Fonseca, Sameh Elnikety, Christina Delimitrou, and Ricardo Bianchini.
\newblock Faster and cheaper serverless computing on harvested resources.
\newblock In {\em Proceedings of the ACM SIGOPS 28th Symposium on Operating Systems Principles}, pages 724--739, 2021.

\bibitem{zhao2024lora}
Justin Zhao, Timothy Wang, Wael Abid, Geoffrey Angus, Arnav Garg, Jeffery Kinnison, Alex Sherstinsky, Piero Molino, Travis Addair, and Devvret Rishi.
\newblock Lora land: 310 fine-tuned llms that rival gpt-4, a technical report.
\newblock {\em arXiv preprint arXiv:2405.00732}, 2024.

\bibitem{zheng2023judging}
Lianmin Zheng, Wei-Lin Chiang, Ying Sheng, Siyuan Zhuang, Zhanghao Wu, Yonghao Zhuang, Zi~Lin, Zhuohan Li, Dacheng Li, Eric Xing, et~al.
\newblock Judging llm-as-a-judge with mt-bench and chatbot arena.
\newblock {\em Advances in Neural Information Processing Systems}, 36:46595--46623, 2023.

\bibitem{zheng2023efficiently}
Lianmin Zheng, Liangsheng Yin, Zhiqiang Xie, Jeff Huang, Chuyue Sun, Cody Hao~Yu, Shiyi Cao, Christos Kozyrakis, Ion Stoica, Joseph~E Gonzalez, et~al.
\newblock Efficiently programming large language models using sglang.
\newblock {\em arXiv e-prints}, pages arXiv--2312, 2023.

\bibitem{zhong2024distserve}
Yinmin Zhong, Shengyu Liu, Junda Chen, Jianbo Hu, Yibo Zhu, Xuanzhe Liu, Xin Jin, and Hao Zhang.
\newblock Distserve: Disaggregating prefill and decoding for goodput-optimized large language model serving.
\newblock {\em arXiv preprint arXiv:2401.09670}, 2024.

\bibitem{zhu2023multilingual}
Wenhao Zhu, Hongyi Liu, Qingxiu Dong, Jingjing Xu, Shujian Huang, Lingpeng Kong, Jiajun Chen, and Lei Li.
\newblock Multilingual machine translation with large language models: Empirical results and analysis.
\newblock {\em arXiv preprint arXiv:2304.04675}, 2023.

\end{thebibliography}

\end{document}